European Commission

# European Defence Readiness

Kancs, D.

2025

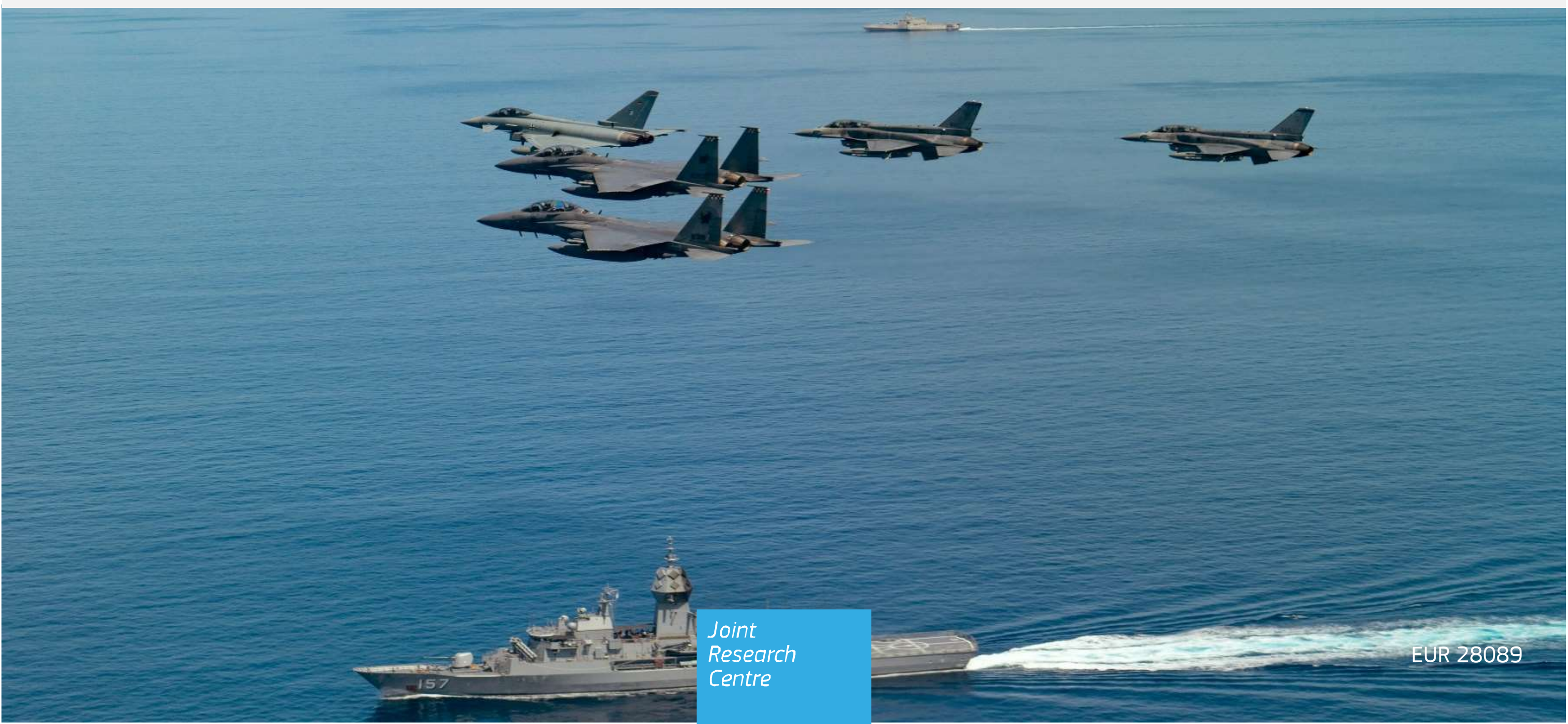

Joint Research Centre

This document is a publication by the Joint Research Centre (JRC), the European Commission's science and knowledge service. It aims to provide evidence-based scientific support to the European policymaking process. The contents of this publication do not necessarily reflect the position or opinion of the European Commission. Neither the European Commission nor any person acting on behalf of the Commission is responsible for the use that might be made of this publication. For information on the methodology and quality underlying the data used in this publication for which the source is neither Eurostat nor other Commission services, users should contact the referenced source. The designations employed and the presentation of material on the maps do not imply the expression of any opinion whatsoever on the part of the European Union concerning the legal status of any country, territory, city or area or of its authorities, or concerning the delimitation of its frontiers or boundaries.

**Contact information**
Name: d'Artis Kancs
Address: Via Enrico Fermi, 21027 Ispra (VA), Italy
Email: d'artis.kancs@ec.europa.eu
https://joint-research-centre.ec.europa.eu/dartis-kancs

# Contents

## Abstract

Europe's defence readiness is being tested by a deteriorating security environment, prolonged war on the continent and growing uncertainty over the future security environment. Against this background, the EU's White Paper for European Defence – Readiness 2030 identifies critical capability gaps, fragmentation of the European defence market and constraints on military mobility. This study assesses two dimensions of European defence readiness empirically: mobilisation readiness and whole-of-society resilience. The current-readiness assessment reveals substantial weaknesses in defence-industrial mobilisation relative to a 1990 readiness benchmark, particularly in production capacity, market fragmentation and security of supply. Whole-of-society resilience is stronger overall but highly heterogeneous across European allies, with the Nordic countries displaying comparatively high levels of preparedness and resilience. To examine how these vulnerabilities evolve under deteriorating external conditions, the study complements the current-readiness assessment with a forward-looking scenario analysis of a hypothetical 'total trade war' involving a complete cessation of trade with China, Russia, Iran and North Korea. The simulations show that such a systemic shock would amplify existing weaknesses in both defence-industrial production and economic resilience. Comparing alternative adjustment trajectories indicates that early and gradual de-risking from vulnerable foreign-input dependencies can substantially reduce adjustment costs relative to abrupt decoupling triggered by geopolitical events. The results therefore suggest that preparedness depends not only on increasing current capabilities, but also on reducing structural vulnerabilities before a systemic shock occurs.

**Keywords**: Preparedness, readiness, Europe, systemic shock, CRINK.

**JEL**: H56, H57, L11.

## Acknowledgements

The authors acknowledge helpful comments from participants of meetings of the Working Party on preparedness, response capability and resilience to future crises, the NATO Science and Technology Organization (STO) Operations Research and Analysis (OR&A) conference in Malaga, NATO Allied Command Transformation (ACT) Experimentation and Wargaming Branch International Concept Development & Experimentation (ICD&E) 2024 Conference in Vilnius, and NATO Disruptive Modelling and Simulation Technologies Symposium of the STO Modelling and Simulation Group in La Spezia. The authors are solely responsible for the content of the report. The views expressed are purely those of the authors and may not in any circumstances be regarded as stating an official position of the European Commission, NATO or Latvian National Armed Forces. Any remaining errors are ours.

# European Defence Readiness[1]

**d'Artis Kancs**

Directorate-General Joint Research Centre, European Commission;
Science Research Innovation Implementation Centre, National Armed Forces

d'artis.kancs@ec.europa.eu

[1] This research is part of a broader EU work on enhancing European Preparedness and Readiness.

# 1 Introduction

The EU's White Paper on European Defence Readiness 2030 identifies critical capability gaps, fragmentation of the defence market and constraints on military mobility, and provides the framework for the ReArm Europe plan (JOIN 2025). It also recognises that European defence readiness must remain robust under worst-case scenarios.

This question has become more urgent as Europe faces increasingly multidimensional and cross-border threats. Three decades of post-Cold War relative peace ended with Russia's full-scale invasion of Ukraine, while security, political, economic and environmental crises have become increasingly interconnected. The traditional peace-war continuum has increasingly given way to an environment in which cooperation, competition and conflict coexist, both within the Alliance and in relations with external actors such as China. Michta (2024) characterises this environment as the early stage of a system-transforming conflict involving a newly consolidated 'axis of dictatorships'. At the same time, political, military and economic differences within the transatlantic Alliance complicate collective responses.

Against this background, the study asks two questions: how prepared is Europe to withstand systemic shocks and protracted conflict, and which strategies can strengthen that readiness? We examine two dimensions of defence readiness in Europe: mobilisation readiness for protracted conflict and whole-of-society resilience. Both are central to a country's capacity to sustain a prolonged crisis (Betts 1995; Monaghan et al. 2024).[2]

The study addresses an important gap in the literature. Quantitative assessments of European defence readiness remain constrained by limited public data, while contemporary research linking investment, capabilities and readiness in the post-2022 security environment remains sparse. Much of the foundational literature on these relationships dates from the Cold War, even though warfare has evolved and European military capacity has declined substantially since then. We therefore combine public and confidential defence-production data with NATO's SFA23 and FOE24 assessments and examine them through a current-readiness assessment and a forward-looking scenario exercise.

The current-readiness assessment benchmarks mobilisation readiness and whole-of-society resilience using historical and contemporary European data. It identifies three structural weaknesses in defence-industrial mobilisation: insufficient production capacity, underused potential in the European defence market and vulnerabilities in security of supply. Whole-of-society resilience is more heterogeneous: Nordic allies perform strongly, including by global standards, whereas weaknesses in several southern allies create vulnerabilities for the Alliance as a whole.

The scenario analysis then asks how these weaknesses evolve when external conditions deteriorate. Using defence-industrial production and economic resilience as examples, we simulate a 'total trade war' involving a complete cessation of trade with China, Russia, Iran and North Korea (CRINK), drawing on NATO's SFA23 and FOE24 projections. We use an empirically validated mathematical model to assess the effects on European strategic readiness. The results indicate sizeable losses in industrial production and economic resilience in the absence of timely policy action. They also show that gradual de-risking from vulnerable foreign-input dependencies is less costly than adjustment forced by an abrupt geopolitical shock.[3]

[2] For simplicity, the terms 'European allies' and 'Europe' refer throughout to the 28 European NATO members included in the analysis. References to the Alliance concern this European subset unless otherwise stated.

[3] https://web.jrc.ec.europa.eu/policy-model-inventory/explore/models/model-eu-ems/

The report proceeds as follows. Section 2 assesses Europe's current readiness, first setting out the strategic-readiness framework and then examining mobilisation readiness and whole-of-society resilience. Section 3 stress-tests these findings under a severe external shock and compares alternative policy trajectories. Section 4 draws out implications for strategic decision-makers, and Section 5 concludes.

# 2 Assessment of current readiness

Recent crises, wars and renewed great-power competition raise a basic question: is Europe prepared not only for a major crisis or war, but also for one that is prolonged? Existing readiness metrics often capture preparedness at a point in time and therefore provide only a partial answer (Galvin 2022). They are less well suited to assessing whether countries can mobilise industry, personnel and materiel over an extended period or sustain whole-of-society resilience under prolonged pressure (Monaghan et al. 2024). This section addresses these blind spots through a current-readiness assessment of selected dimensions of strategic readiness.

Definition: "*European Defence Readiness can be defined as a steady state of preparedness of the Union and its member states to protect the security of its citizens, the integrity of its territory and critical assets or infrastructures, and its core democratic values and processes. This includes an ability to provide military assistance to its partners, such as Ukraine.*" (JOIN 2024)[4]

## 2.1 Strategic readiness framework

We use the U.S. Department of Defense Strategic Readiness Framework (SRF) as an organising framework for the current-readiness assessment (US DoD 2023). Its value for this study is not the enumeration of readiness dimensions in itself, but the distinction it enables between meeting current demands and building the capability and capacity required for future crises or conflict. The framework also emphasises data-driven assessment of trade-offs, risks and opportunities associated with strategic choices.

The SRF groups preparedness into ten dimensions: sustainment, modernisation, allies and partners, business systems and organisational effectiveness, human capital, global posture, force structure, resilience, operational readiness and mobilisation (Watts et al. 2024). For Europe, this provides a coherent structure for identifying areas of strength and weakness and the policy levers that can alter the balance.

Given the study's scope, we focus on two dimensions: mobilisation readiness and whole-of-society resilience. This choice is both conceptual and empirical. Betts (1995) distinguishes operational, structural and mobilisation readiness, while Monaghan et al. (2024) identify defence industry, defence capabilities and resilience as particularly weak areas in Europe. Operational readiness, including force mobility, lies beyond the scope of this report. The analysis forms part of a broader EU work stream on preparedness, response capacity and resilience that covers dimensions not examined here.

## 2.2 Mobilisation readiness in Europe

### 2.2.1 Capability baseline and readiness benchmark

Mobilisation readiness concerns the ability to convert available civilian and military resources into additional military capacity. It encompasses industry, personnel and materiel (Campbell 1952). Our analysis focuses on industry: the capacity to expand production, develop additional industrial capabilities and provide the equipment and services required to sustain the force. In this sense, mobilisation readiness depends on both convertibility and expansibility - the ability to organise and scale national resources for a prolonged crisis or war effort (Betts 1995).

Campbell (1952) distinguishes three stages in the transition from a peacetime to a war-ready economy. The first, the 'rearmament' or 'mobilisation hump', shifts the economy towards higher military output and expands relevant productive capacity. The second, 'expansion', adds personnel, reserves and production facilities until the country reaches a steady state of mobilisation readiness. The third, 'total mobilisation', redirects the wider economy towards the war effort. The European Defence Readiness objective is closest to the second stage: a steady state in which sufficient industrial, personnel and materiel capacity is available before a full-scale war begins.

[4] https://eur-lex.europa.eu/legal-content/EN/TXT/?uri=CELEX:52024JC0010

European political commitments point in this direction. The 2025 Special European Council on European Defence and Ukraine called for faster mobilisation of instruments and financing, stronger defence readiness, reduced strategic dependencies and a stronger European defence technological and industrial base (EUCO 6/25). The central analytical problem is therefore not the absence of political ambition, but its conversion into military capability and, ultimately, into security outcomes (Becker and Bell 2020).

Europe nevertheless lacks a common set of measurable readiness targets against which progress can be assessed. The contrast with the United States after the outbreak of the Korean War is instructive. In 1951, the U.S. Director of Defence Mobilisation specified explicit mobilisation objectives, including the capacity to produce equipment for national forces, allies and reserves sufficient for the first year of a full-scale war. Europe has a definition of Defence Readiness, but no equivalent quantitative benchmark.

To construct a measurable proxy for European Defence Readiness, we use the military capabilities of European allies at the end of the Cold War as a historical benchmark. Drawing on IISS (1991), we calculate 1990 inventories for eight categories of major weapon systems across ground, naval and air forces: main battle tanks, infantry fighting vehicles, armoured personnel carriers, artillery, mortars, submarines, principal surface combatants and combat aircraft. We then compare these inventories with the same categories in 2024 using IISS (2025). Table A1 reports the underlying stocks, while Table 1 expresses current holdings as a percentage of their 1990 level for twelve European allies.

**Table 1**: Share of stocks of key weapon systems of selected European allies in 2024 relative to 1990 (%)

| | BEL | DNK | FRA | GER | ITA | NLD | NOR | PRT | ESP | GBR | FIN | SWE |
|---|---|---|---|---|---|---|---|---|---|---|---|---|
| Personnel | 27.5 | 44.6 | 44.6 | 37.8 | 44.8 | 33.2 | 77.7 | 65.0 | 47.5 | 47.0 | 87.4 | 23.6 |
| MBT | 0.0 | 8.8 | 14.8 | 4.5 | 12.3 | 0.0 | 17.1 | 23.3 | 32.7 | 16.2 | 166.7 | 14.0 |
| IFV | 5.2 | na | 41.4 | 13.4 | na | 11.9 | 171.7 | 18.1 | na | 17.6 | 294.4 | na |
| APC | 5.5 | 65.5 | 69.6 | 7.8 | 9.5 | 9.0 | 260.0 | 146.3 | 52.5 | 24.4 | 223.3 | 140.8 |
| ARTY/HOW | 2.3 | 0.2 | 3.3 | 1.5 | 5.6 | 1.6 | 6.0 | 39.9 | 16.3 | 11.6 | 118.2 | 2.5 |
| MOR | 10.6 | 2.7 | 10.7 | 7.7 | 26.7 | 29.8 | 114.4 | 148.1 | 71.4 | 72.0 | 47.9 | 15.2 |
| Submarines | | 0.0 | 52.9 | 25.0 | 88.9 | 60.0 | 54.5 | 66.7 | 25.0 | 41.7 | | 33.3 |
| PSC | 50.0 | 166.7 | 53.7 | 78.6 | 56.3 | 33.3 | 80.0 | 40.0 | 55.0 | 33.3 | | |
| Aircraft | 27.0 | 46.2 | 29.8 | 29.9 | 39.3 | 18.4 | 52.1 | 43.4 | 63.8 | 24.4 | 75.4 | 21.0 |

*Source: Authors' calculations based on IISS (1991, 2025). Notes: Current holdings are expressed relative to 1990 stocks (1990 = 100). Underlying unit counts are reported in Table A1; "na" denotes data not available.*

Table 1 shows a substantial contraction in major weapon inventories since 1990, particularly in land forces. Main battle tank stocks have fallen sharply in every analysed country except Finland; Belgium and the Netherlands no longer field main battle tanks. France, Germany and the United Kingdom retain only 12–16 per cent of their 1990 stocks. Similar declines are visible in infantry fighting vehicles and armoured personnel carriers, while artillery inventories have fallen even more sharply. Naval inventories have also declined, although generally less dramatically, and combat-aircraft stocks range from 18 per cent of their 1990 level in the Netherlands to 75 per cent in Finland.

The scale of these reductions matters because the requirements of a high-intensity conflict remain large. Burilkov and Wolff (2025) estimate that preventing a rapid Russian breakthrough in the Baltic states alone would require at least 1,400 tanks, 2,000 infantry fighting vehicles and 700 artillery systems. These requirements exceed the combined current holdings of the twelve European allies examined here in several categories (Table A1). Munitions depth is equally important: the same authors estimate that one million 155 mm shells would be a minimum requirement for 90 days of high-intensity combat.

**Figure 1**: Problems, drivers and consequences of defence-industrial readiness in Europe

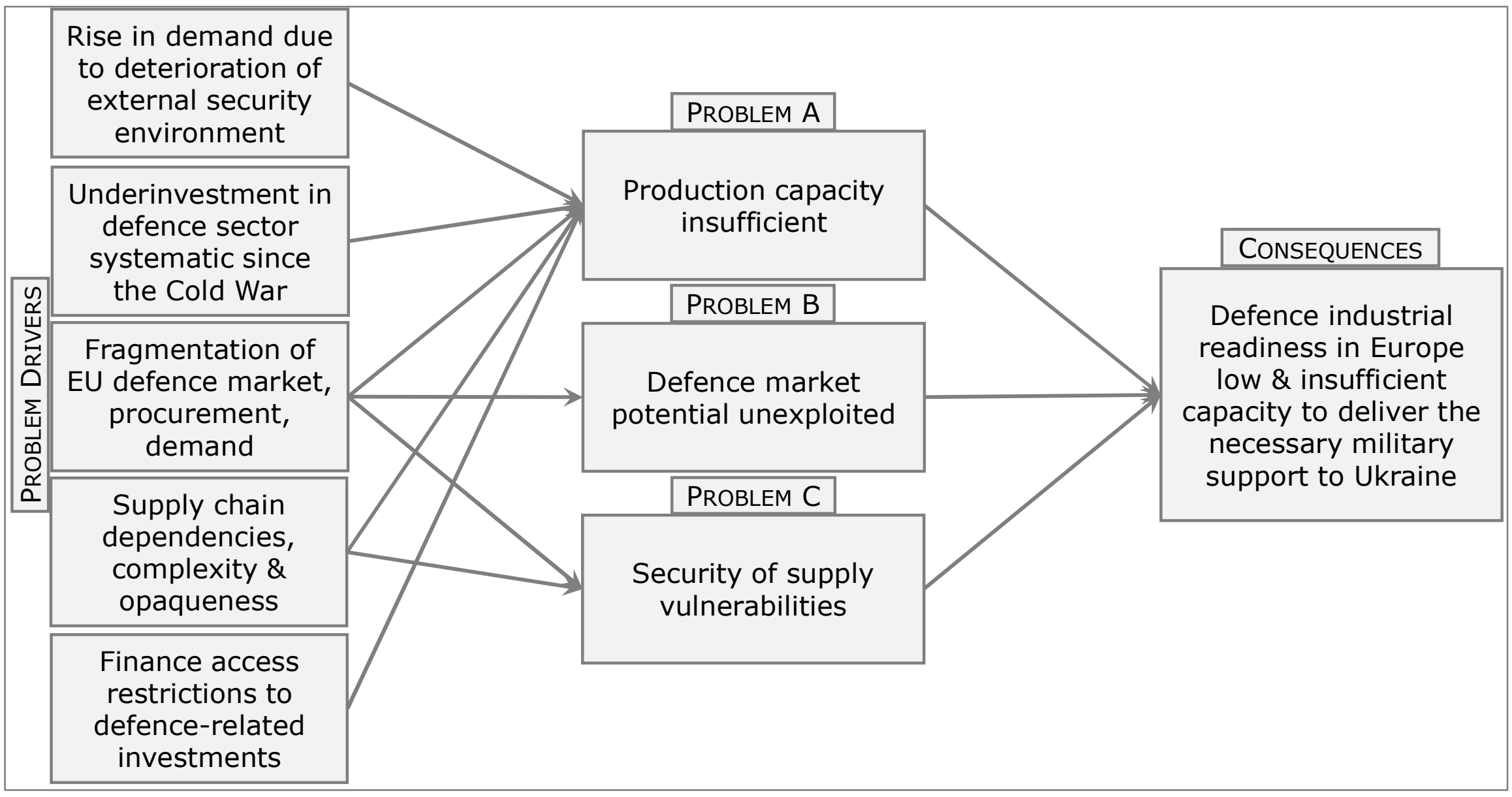

*Source: Authors' presentation based on EC (2024). Notes: The figure summarises the drivers, constraints and consequences developed in Section 2.2.*

The resulting gap is therefore not simply between current and historical stock levels, but between political ambition and the capacity required to sustain credible defence. Figure 1 organises the industrial component of this gap around three linked constraints: whether industry can regenerate and expand stocks, whether fragmented demand prevents Europe from exploiting economies of scale, and whether external dependencies can interrupt production when it is most needed.

### 2.2.2 Industrial regeneration capacity

The first constraint is production capacity (Figure 1). To gauge how quickly the European industrial base could regenerate capability, we estimate the time required to expand 2024 inventories to their 1990 benchmark levels. This inventory-expansion time serves as a proxy for the capacity of the defence industrial base to meet the demands of prolonged conflict. Where 2024 inventories already exceed 1990 levels, the expansion requirement is set to zero.

We follow the methodology of Cancian et al. (2020), which allows us to compute the defence industrial capacity for increasing or replacing existing stockpiles. The inventory expansion time, $Inventory\ Expansion_w$, in years is calculated as:

$$Inventory\ Expansion\ _w = \frac{Inventory\ Objective_w}{Production\ Rate_w} + Production\ Leadtime_w$$

where $Inventory\ Objective_w$ denotes weapon system *w*'s inventory objective which corresponds to 1990 inventories, $Production\ Rate_w$ is the industrial production rate and $Production\ Leadtime_w$ denotes production lead time. The economic production rate is defined as the most efficient peacetime production rate achievable with existing plant capacity and tooling, assuming one eight-hour shift per day, five days per week. The maximum production rate is the highest output achievable with existing tooling when the number of shifts is increased to the maximum feasible level.

Estimating expansion time requires both inventory data and production rates. We combine IISS (2025) inventory data with information on European defence manufacturers from SIPRI (2025). Because comparable European system-level production-rate estimates are unavailable, we use U.S. industrial mobilisation data from Cancian et al. (2020) as a proxy. Since U.S. procurement spending has historically exceeded that of European allies, these estimates are best interpreted as a lower bound on European expansion times under current production conditions (Lucas et al. 2023).

**Table 2**: Average expansion time (years) of current (2024) inventories to 1990 levels in Europe and threshold attrition rate (per cent)

| | Current stocks 2024/1990, % | Production expansion time, y | | Threshold attrition rate, % | |
|---|---|---|---|---|---|
| | | Economic | Maximum | Economic | Maximum |
| MBT | 10.7 | 23.8 | 15.7 | 6.3 | 9.5 |
| IFV | 34.0 | 12.3 | 7.4 | 8.9 | 14.9 |
| APC | 30.7 | 13.0 | 8.2 | 8.4 | 13.2 |
| ARTY/HOW | 8.1 | 15.9 | 9.6 | 9.6 | 16.0 |
| MOR | 33.8 | 8.8 | 5.3 | 12.5 | 20.8 |
| Submarines | 39.3 | 21.1 | 18.2 | 5.0 | 5.8 |
| Principal surface comb | 53.1 | 22.4 | 19.3 | 3.6 | 4.2 |
| Aircraft, combat | 32.5 | 16.5 | 11.0 | 7.7 | 11.6 |

*Source: Authors' computations based on IISS (2025), Cancian et al. (2020) and SIPRI (2025). Notes: Expansion times are reported for economic and maximum production rates. Threshold attrition rates follow the methodology described in the text and Stoll (1990).*

Table 2 shows that restoring 2024 inventories to 1990 levels would take many years even at maximum production rates. Ground systems such as infantry fighting vehicles, armoured personnel carriers, artillery and mortars can generally be expanded faster than complex naval and air systems. Submarines, principal surface combatants and combat aircraft have particularly long expansion times because their production is capital-intensive, specialised and difficult to scale rapidly.

Table 2 also reports threshold attrition rates: the share of inventory that can be lost in each period before replacement capacity becomes insufficient to maintain a zero withdrawal rate. Following Stoll (1990), the measure links industrial regeneration capacity to battlefield losses. Under the Defence Readiness concept used here, once attrition exceeds this threshold, the force can no longer replace losses fast enough to maintain its position.

The war in Ukraine illustrates the scale of the challenge. Oryx (2025) documents more than 20,000 Russian equipment losses and about 7,600 Ukrainian losses during the first three years of the war, including thousands of armoured vehicles. Because Oryx counts only visually confirmed losses, these figures are lower bounds. Even so, the implied attrition exceeds the replacement thresholds estimated in Table 2 for several categories, highlighting the gap between European industrial capacity and the demands of sustained high-intensity warfare.

Production capacity is therefore a structural mobilisation constraint, not simply a short-term shortage. Most national preparedness strategies were designed primarily for threats below the level of war, such as terrorism, natural disasters, cyberattacks or critical-infrastructure failures (Galvin 2022). Although many associated capabilities remain useful in wartime, prolonged conflict would require them to expand substantially, intensifying competition for labour, raw materials, production capacity and logistics (Campbell 1952).

Five factors help explain this constraint. First, the security environment has shifted faster than industrial capacity. Second, decades of underinvestment have reduced productive depth. Third, fragmented national demand has encouraged industry to serve relatively small domestic markets. Fourth, supply-chain bottlenecks limit expansion. Fifth, restricted access to finance can delay or prevent the investment required to increase output (EC 2024). These constraints connect production capacity directly to the market and supply-security problems examined next.

### 2.2.3 Defence market fragmentation

The second constraint is the fragmented structure of the European defence market (Figure 1). Defence demand remains predominantly national, and procurement decisions are often shaped by a systematic preference for domestic suppliers (Olsson 2021). The result is a collection of relatively small national markets rather than a genuinely integrated European market. Because governments are the principal buyers of defence products, fragmented demand is mirrored in fragmented industrial supply.

We quantify market fragmentation using the reciprocal of the Herfindahl-Hirschman Index (HHI) for each weapon system, following Allen et al. (2021). Higher values indicate that inventories are distributed across more manufacturers, while values closer to one indicate greater concentration. The index is expressed as follows:[5]

$$Fragmentation_w = \frac{1}{\sum_w \left( \frac{Number\ of\ equipment\ pieces_{k,w}}{\sum_{j,w} Number\ of\ equipment\ pieces_{j,w}} \right)^2}$$

where  denotes the number of equipment pieces of manufacturer k of weapon system w in European inventories in 2024,  represents the total number of equipment of weapon system w in European inventories in 2024, and the denominator term in brackets is the share of manufacturer k of weapon system w in European inventories in 2024.  is bounded below by one. For weapon systems in which a large share of European stocks comes from one or a few manufacturers, the proxy takes values closer to one; where stocks are distributed across many manufacturers, the proxy takes higher values.$Number\ of\ equipment\ pieces_{k,w} Number\ of\ equipment\ pieces_{j,w} Fragmentation_w$

We measure defence-market fragmentation using SIPRI (2025) data on defence manufacturing. As in Section 2.2.1, we focus on the same categories of ground, naval and air systems, while IISS (2025) provides the corresponding 2024 inventories. Table 3 reports the resulting fragmentation index, the number of weapon types in Europe and the United States, and the market share of the largest European manufacturer.

**Table 3**: European defence market fragmentation in 2024

| | Fragmentation Index | Weapon types Europe Count | Weapon types US Count | Europe largest manufacturer Share |
|---|---|---|---|---|
| MBT | 9.3 | 19 | 1 | 0.28 |
| IFV | 21.9 | 23 | 3 | 0.09 |
| Howitzer | 21.3 | 28 | 2 | 0.14 |
| Combat aircraft | 16.5 | 20 | 7 | 0.15 |
| Principal surface combatants | 33.1 | 41 | 6 | 0.10 |
| Submarines | 4.6 | 14 | 2 | 0.44 |

*Source: Authors' computations based on SIPRI (2025) and IISS (2025). Notes: The fragmentation index is the reciprocal of the Herfindahl-Hirschman Index; higher values indicate greater fragmentation.*

[5] The fragmentation measure is defined as 1/HHI, a convex transformation of the HHI that can generate relatively large values for highly fragmented markets. As a robustness check, we also compute 1-HHI. The qualitative results are unchanged, so only the 1/HHI measure is reported; the alternative results are available from the authors upon request.

Table 3 confirms substantial variation across weapon systems. Submarines and main battle tanks are relatively concentrated, while infantry fighting vehicles, howitzers and especially principal surface combatants are highly fragmented. The pattern is visible both in the number of manufacturers and in the variety of weapon types in European inventories. This fragmentation limits standardisation and weakens the scale advantages that could otherwise follow from Europe's aggregate demand.

The same pattern appears in cross-border procurement. Although EU defence-equipment expenditure rose by about 65 per cent between 2017 and 2022, intra-EU trade in defence-related products did not increase (Eurostat 2025). The share of intra-EU defence procurement fell from 22 per cent to 15 per cent over the same period, far below the level of integration observed in the wider European economy. Rising demand therefore continues to reinforce nationally segmented supply chains rather than a more integrated European market (EC 2024).

Fragmentation imposes costs through duplication, foregone economies of scale, weak collaborative procurement and R&D, dependence on non-European suppliers and reduced interoperability (Olsson 2021; Centrone and Fernandes 2024). Empirical work suggests that larger production runs can materially reduce unit costs (Hartley 2006; McKinsey 2013). Yet collaborative procurement remained below 30 per cent in 2024, and only 6 per cent of defence R&D spending was collaborative (EDA 2024). In parallel, 78 per cent of EU defence-procurement expenditure in 2022-2023 went to suppliers outside Europe (EC 2024). The central implication is that fragmented demand reduces the amount of capability Europe can obtain from a given level of spending.

### 2.2.4 Security of supply

The third constraint is security of supply (Figure 1). External dependencies may be manageable in peacetime but become critical during a major crisis or war, when export controls, transport disruption, higher demand and the deliberate weaponisation of supply chains can restrict access to intermediate inputs. Exposure is particularly acute in critical raw materials: China supplies a substantial share of inputs used by the European defence sector and has already imposed export restrictions on several strategic materials (EC 2024).

To identify vulnerabilities arising from the European defence industry's foreign dependencies, we distinguish between supply-side dependence and demand-side exposure. Following Baldwin et al. (2023), we compute the defence sector's sourcing-side exposure, Foreign Input Reliance (FIR), as:

$$FIR_{c,j} = \sum_{c'=1}^{N} \sum_{j'}^{J} \frac{Foreign\ output_{c,j}^{c',j'}}{Gross\ output_{c,j}}$$

and the defence sector's reliance on foreign markets on the sales side (Foreign Market Reliance (FMR)) as:

$$FMR_{c,j} = \sum_{c'=1}^{N} \sum_{j'}^{J} \frac{Domestic\ output\ in\ partner\ gross\ output_{c,j}^{c',j'}}{Gross\ output_{c,j}}$$

where subscripts $c$ and $j$ denote, respectively, countries and industries; $Foreign\ output_{c,j}^{c',j'}$ denotes gross output of industry $j'$ in country $c'$ used in the production of industry $j$ in country $c$; and $Domestic\ output\ in\ partner\ gross\ output_{c,j}^{c',j'}$ denotes gross output of industry $j$ in country $c$ to industry $j'$ in country $c'$.

Because European defence production is embedded in global value chains (Kancs 2024), FIR and FMR account for both direct and indirect trade links using Leontief flows from inter-country input-output tables. The Leontief inverse captures inputs embodied in goods that pass through third countries, not only direct bilateral imports. FIR and FMR therefore measure ultimate exposure across higher-tier suppliers and buyers rather than only first-order trading relationships.[6]

We estimate FIR and FMR using OECD Inter-Country Input-Output (ICIO) data, complemented by defence-manufacturing information and Eurostat External Trade Statistics (Comext). Defence-related activities are identified using two NACE Rev. 2 four-digit industries - manufacture of weapons and ammunition, and manufacture of military fighting vehicles - which are aggregated into a single defence-industry sector. Because systemic disruptions can interrupt an entire shipment rather than only the value added in the affected country, FIR and FMR are calculated on gross output. The underlying defence data are accessed under a special agreement and can therefore be reported only in aggregated form.[7][8]

**Table 4**: European defence industry's Foreign Input Reliance (FIR, %) in 2024

| FIR | BEL | DNK | FRA | DEU | ITA | NLD | NOR | PRT | ESP | UK | FIN | SWE | CHN |
|---|---|---|---|---|---|---|---|---|---|---|---|---|---|
| BEL | | 4.6 | 5.4 | 4.1 | 4.5 | 7.0 | 5.4 | 6.3 | 6.3 | 4.7 | 6.5 | 4.5 | 11.6 |
| DNK | 4.8 | | 3.5 | 3.8 | 3.9 | 5.9 | 3.6 | 3.9 | 5.5 | 5.1 | 4.2 | 3.4 | 8.8 |
| FRA | 9.4 | 5.6 | | 14.4 | 7.2 | 9.0 | 6.6 | 6.4 | 9.9 | 5.2 | 6.2 | 9.9 | 9.2 |
| DEU | 5.5 | 7.7 | 6.4 | | 4.9 | 5.6 | 5.1 | 5.1 | 5.9 | 4.9 | 7.7 | 4.7 | 13.1 |
| ITA | 7.5 | 6.0 | 8.5 | 13.9 | | 5.5 | 5.9 | 6.9 | 8.3 | 4.0 | 7.4 | 9.6 | 11.4 |
| NLD | 5.2 | 6.5 | 6.6 | 5.1 | 6.6 | | 4.0 | 6.4 | 4.6 | 3.9 | 4.2 | 4.9 | 10.1 |
| NOR | 4.8 | 3.0 | 2.8 | 3.0 | 3.6 | 4.0 | | 4.0 | 3.0 | 2.9 | 5.1 | 2.9 | 7.9 |
| PRT | 6.0 | 4.2 | 7.1 | 5.5 | 5.8 | 5.6 | 5.9 | | 4.9 | 6.8 | 6.0 | 4.0 | 11.4 |
| ESP | 4.5 | 4.5 | 5.6 | 6.3 | 5.2 | 3.8 | 4.0 | 4.7 | | 4.0 | 5.2 | 5.6 | 9.0 |
| UK | 6.1 | 5.9 | 6.7 | 9.1 | 3.8 | 6.0 | 5.0 | 4.8 | 6.5 | | 5.5 | 6.1 | 12.4 |
| FIN | 4.4 | 3.7 | 3.1 | 4.6 | 4.0 | 3.3 | 3.1 | 3.4 | 4.9 | 3.7 | | 2.9 | 7.1 |
| SWE | 4.6 | 3.4 | 4.4 | 3.9 | 3.3 | 3.9 | 4.3 | 4.9 | 4.4 | 4.1 | 3.8 | | 8.2 |
| CHN | 1.1 | 0.8 | 1.2 | 2.5 | 0.7 | 2.4 | 0.8 | 0.8 | 1.3 | 0.8 | 0.7 | 1.2 | |

*Source: Authors' estimations based on ICIO, EUregio and Comext data. Notes: FIR captures both direct and indirect foreign-input exposure through Leontief production linkages; methodological details are provided in the text.*

Table 4 shows that the European defence industry's average reliance on intermediate inputs from China is 9.9 per cent, with substantial variation across countries: from 7.1 per cent in Finland to 13.1 per cent in Germany. Dependence also varies across product lines within the aggregated defence sector, although these differences cannot be identified in our data. Nitrocellulose, the main ingredient of gunpowder, is a particularly stark example: China supplies 78.5 per cent of the nitrocellulose used by European defence manufacturers and is also the world's largest exporter (Eurostat 2025). Scaling up European explosives production therefore depends not only on domestic capacity but also on uncertain future access to Chinese inputs. The asymmetry is pronounced: the bottom row of Table 4 shows that Chinese defence manufacturing relies far less on European inputs; for example, only 1.2 per cent of its intermediate inputs are sourced from France.

European dependence on Chinese inputs reflects China's wider position in global intermediate-goods trade. Although China accounts for a relatively modest 6.6 per cent of global arms exports (SIPRI 2025), it represented almost 21 per cent of global manufacturing trade in 2024 according to Comext data. The strategic issue is therefore not only who supplies complete weapon systems, but who supplies the components and materials needed to produce them. As former NATO Secretary General Stoltenberg warned in 2024, "Russia used gas as a weapon to try to coerce us. We must not make the same mistake with China." Although security of supply remains a national competence in the EU, increasingly cross-border and global supply chains give it a clear European dimension (Kancs 2024). Given the rising cost and complexity of advanced defence capabilities, no single European country can realistically develop, produce and sustain the full spectrum of capabilities on a purely national basis.

---

[6] The Leontief approach captures both direct and indirect import dependence. Unlike customs data, which record bilateral imports directly, input-output linkages also capture intermediate inputs originating in one country that reach the destination after being embodied in goods produced in third countries (Baldwin et al. 2023).

[7] http://oe.cd/icio

[8] https://ec.europa.eu/eurostat/comext/

Two factors reinforce these security-of-supply vulnerabilities (Figure 1). First, fragmented national markets make coordinated sourcing and stockpiling more difficult, particularly during major crises or war. The COVID-19 response showed the value of joint action: the EU established a clearing house for medical equipment, temporarily waived customs duties and VAT on selected imports, and created common reserves through rescEU. Second, limited visibility over complex, multi-tier defence supply chains makes it difficult to identify dependencies on third countries for critical components. Because production networks are both deep and highly interconnected, governments and firms may not know the origin of components several tiers removed from the prime contractor. This information gap itself constitutes a vulnerability that cannot be addressed effectively at national level alone.

## 2.3 Whole-of-society resilience in Europe

Resilience is the ability to maintain essential functions and services under adverse conditions, including when warning is limited or absent (Ottosson et al. 2024). Whole-of-society resilience requires stockpiles, facilities and infrastructure for mobilisation, robust production and distribution systems, and the organisational flexibility to reallocate resources as demands change during a prolonged conflict. Ukraine's experience illustrates why these capabilities matter: as casualties accumulate, equipment is damaged or destroyed and supply routes are disrupted, societies must regenerate capacity while continuing to support military operations and essential civilian functions (NATO 2024).

**Figure 2**: Problems, drivers and consequences of whole-of-society readiness in Europe

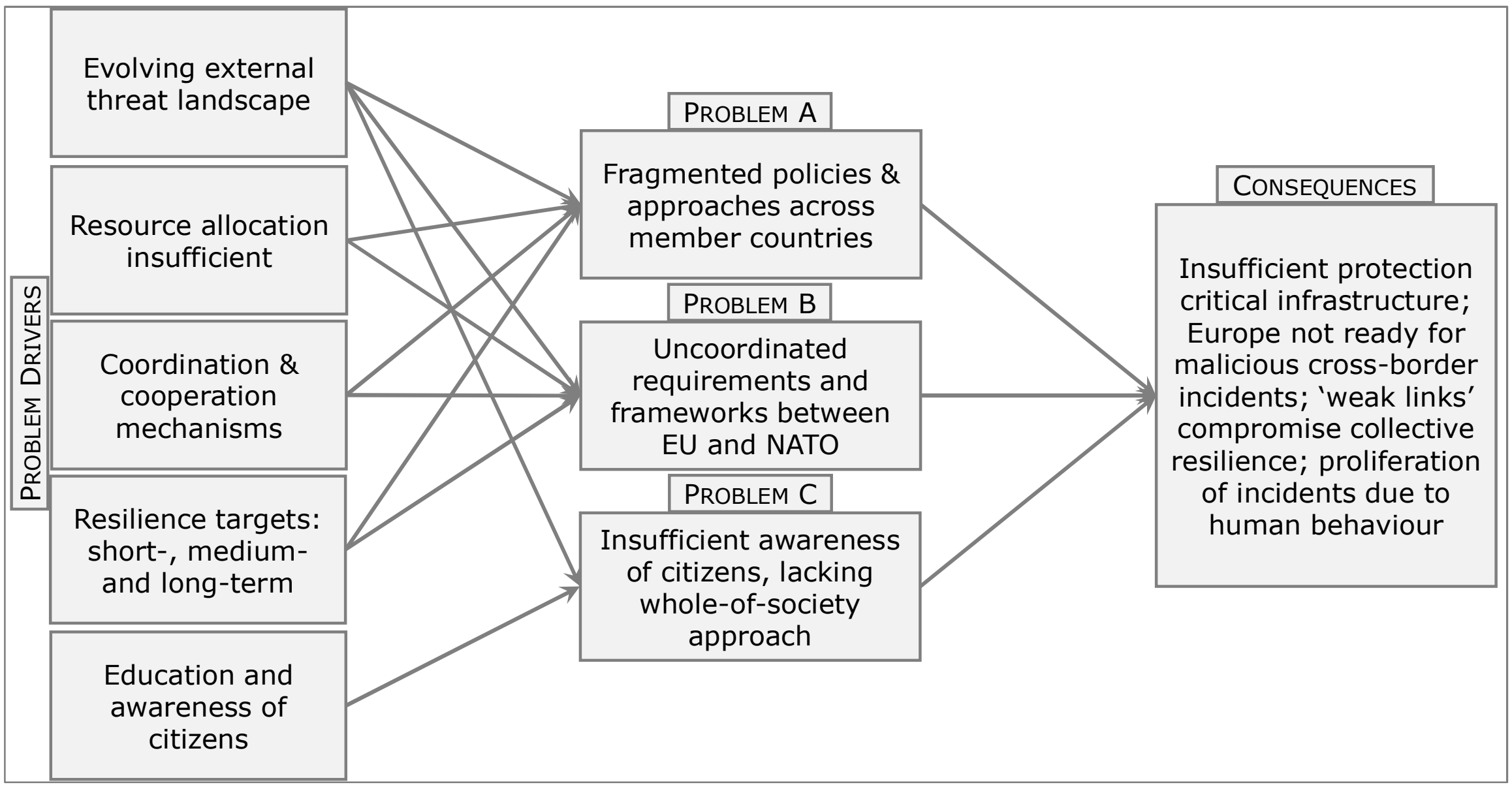

*Source: Authors' presentation based on EC (2025). Notes: The figure summarises the drivers, constraints and consequences developed in Section 2.3.*

To assess Europe's whole-of-society resilience, we apply the same structural approach used for defence-industrial readiness and identify the principal problems, drivers and consequences of whole-of-society resilience. Figure 2 highlights three recurring weaknesses: fragmented national policies and approaches; uncoordinated EU and NATO requirements and frameworks; and insufficient public awareness and whole-of-society preparedness. Together, these weaknesses create a gap between resilience ambitions and practical capacity, including shortcomings in critical-infrastructure protection, vulnerability to cross-border disruption and the risk that weaker national systems reduce collective resilience.

Several approaches to assessing whole-of-society resilience have been developed in the literature (Nederveen et al. 2024). We use the Enhanced Analytic Resilience Index (EARI) proposed by E-ARC (2024), which covers five components: prerequisites of resilience, preparedness, shock resistance, crisis recovery and risk exposure. The prerequisites component captures institutional, social, economic and human-capital conditions that support resilience. Preparedness measures individual and institutional capabilities, state capacity, health security, investment capacity and related socio-economic factors. Shock resistance covers continuity of government, energy and transport resilience, civil communications, food and water security, vulnerable populations and health-system capacity. Crisis recovery captures adaptive capacity, health-security financing and recent societal shocks. The fifth component, exogenous risk exposure, measures factors such as climate and seismic risk. Because such exposure can change more rapidly than the data needed to measure it reliably, we exclude this component from the empirical analysis.

$$\overline{resilience}_q = \frac{1}{m_q} \sum_{recorded} resilience_{q,c}$$

where the observed value is the individual resilience variable for indicator q in country c, and m is the number of recorded, non-missing observations. Scores are normalised to a scale from 0 to 10. A score of zero represents a hypothetical case in which a systemic shock causes persistent damage with no recovery, while a score of 10 represents full resilience: the country absorbs the shock without disrupting the continuity of Alliance activities and immediately returns to its pre-shock steady state.$resilience_{q,c} R_{q,c}$

We assess whole-of-society resilience across European allies using 2024 data from the Corruption Perceptions Index (Transparency International), World Risk Index (UNFCCC), Fragile States Index and State Resilience Index (Fund for Peace), Global Health Security Index (Johns Hopkins Center for Health Security), Global Resilience Index (FM), Global Freight Resilience Index (WS), World Development Indicators (World Bank) and Climate Change Indicators (IMF). Combining these sources in the EARI composite index provides a measure of national and collective capacity to prepare for, resist, respond to and recover from shocks while maintaining the continuity of Alliance activities.

Table 5 summarises whole-of-society resilience in 2024 for twelve European allies. Mean scores range from 6.7 for the prerequisites of resilience to 7.9 for shock resistance, placing overall resilience in the upper part of the E-ARC scale. To aid interpretation, the final column reports each country's global E-ARC ranking for 2023 (E-ARC 2024). The United States ranks 17th globally and is not shown in the table. In comparative terms, whole-of-society resilience therefore appears less vulnerable than the defence-industrial mobilisation dimension assessed in Section 2.2.

The estimates reveal substantial heterogeneity across both countries and resilience dimensions. Nordic countries - Finland, Norway, Denmark and Sweden - rank highest, including in global comparison, while Italy, Spain and Portugal generally score lower. The prerequisites-of-resilience component averages 6.69, ranging from 6.11 in the United Kingdom and 6.15 in Italy to above 8.0 in several Nordic countries. Preparedness averages 7.20, from 6.84 in Spain to 8.58 in Norway. Shock resistance is the strongest component, averaging 7.95 and exceeding 9.0 in Finland, Norway and the Netherlands. Crisis recovery, adaptation and post-shock thriving average 7.47; Italy records the lowest score at 6.64, whereas all four Nordic countries and the United Kingdom score above 8.0. Overall, European resilience is uneven rather than uniformly weak, and this heterogeneity matters for collective readiness as much as national averages do.

Two structural weaknesses identified in Figure 2 - fragmented national approaches and uncoordinated or non-binding EU and NATO frameworks - help explain the variation in whole-of-society resilience across European allies. These differences matter for the Alliance's collective ability to prepare for, resist, respond to and recover from shocks. They are especially important for cross-border infrastructure in energy, finance, data, telecommunications and transport, where vulnerabilities in one country can have Alliance-wide consequences (Polyakova et al. 2024).

**Table 5:** Enhanced Analytic Resilience Index for selected European countries, 2024

| | Pre-requisites | Pre-paradness | Shock resistance | Crisis recovery | Global rank E-ARC 2023 |
|---|---|---|---|---|---|
| BEL | 6.56 | 7.61 | 8.40 | 7.50 | 20 |
| DNK | 7.96 | 8.42 | 8.50 | 8.35 | 4 |
| FRA | 6.69 | 7.42 | 8.01 | 7.48 | 22 |
| DEU | 7.18 | 7.55 | 8.72 | 7.77 | 13 |
| ITA | 6.15 | 7.27 | 7.95 | 6.64 | 29 |
| NLD | 7.35 | 7.57 | 9.02 | 8.33 | 10 |
| NOR | 7.84 | 8.58 | 9.22 | 8.05 | 3 |
| PRT | 6.64 | 7.48 | 7.88 | 7.76 | 21 |
| ESP | 6.57 | 6.84 | 7.65 | 7.59 | 24 |
| UK | 6.11 | 7.62 | 7.97 | 8.53 | 19 |
| FIN | 7.93 | 8.18 | 9.39 | 8.54 | 1 |
| SWE | 8.15 | 8.18 | 8.00 | 8.56 | 6 |

*Source: Authors' computations based on E-ARC (2024) methodology. Notes: Scores are normalised on a 0–10 scale. The exogenous risk-exposure component is excluded from the empirical analysis; the global rank refers to 2023.*

A third weakness - limited public awareness and an incomplete whole-of-society approach - is also unevenly distributed. Several Nordic countries, which score highly on whole-of-society resilience, have institutionalised broad societal participation in preparedness. Finland's 'comprehensive security' model coordinates government authorities, businesses, NGOs and citizens around vital societal functions, including defence capability, internal security, infrastructure, security of supply and psychological resilience. Sweden's 'total defence' model similarly integrates military and civil defence and treats physical, psychological and economic resilience as mutually reinforcing. Since 1943, the Swedish government has periodically issued crisis-preparedness guidance to all households through the brochure 'If Crisis or War Comes'.[9][10]

## 2.4 Synthesis: readiness under current conditions

The current-readiness assessment identifies a common problem across the two readiness dimensions: Europe must be able not only to hold capabilities at a point in time, but also to regenerate military capacity and sustain essential societal functions over a protracted crisis. Mobilisation readiness is constrained by reduced inventories, slow industrial regeneration, market fragmentation and foreign-input exposure, while whole-of-society resilience varies substantially across allies, creating collective vulnerabilities.

---

[9] turvallisuuskomitea.fi/en/comprehensive-security/

[10] government.se/government-policy/total-defence/

# 3 Scenario analysis

Section 2 assesses readiness under current conditions; Section 3 asks how robust that readiness would remain if the external environment deteriorated sharply. The scenario analysis therefore serves as a forward-looking stress test of two weaknesses identified in the current-readiness assessment: defence-industrial dependence on foreign inputs and the economic capacity required to sustain a protracted crisis. Three differences should be kept in mind when interpreting the results. First, the temporal scope is ex ante (2025–2040), whereas Section 2 provides a contemporaneous assessment. Second, the systemic shock considered here is a full-scale trade war, while Section 2 considers readiness for protracted conflict more generally. Third, the shock is modelled at a more aggregated geographic level than the twelve-country assessment in Section 2.

## 3.1 SFA23/FOE24 scenario simulation

The scenario analysis examines how deteriorating external conditions could affect Europe's readiness and how policy choices made in advance could strengthen preparedness. We construct a hypothetical systemic shock from NATO's SFA23 and FOE24 projections: a 'total trade war' involving a complete cessation of trade with CRINK. Using defence-industrial production and economic resilience as examples, we simulate the shock with the EU-EMS model (Kancs 2024) and quantify its effects on European defence-supply vulnerabilities. The model is empirically validated and has previously been used to study preparedness and readiness in Europe.

We use NATO's SFA23 and FOE24 to define a high-impact scenario in which a severe geoeconomic shock tests both mobilisation readiness and whole-of-society resilience. SFA23 identifies key drivers of change and their implications for the Alliance over the next two decades, while FOE24 assesses how evolving actors, technologies and operating conditions may affect the employment of military capabilities. Because future shocks are inherently uncertain, we also draw on Ilut and Schneider's (2023) distinction between risk and ambiguity. Recent events such as the COVID-19 pandemic and Russia's full-scale invasion of Ukraine illustrate how abrupt shocks can alter strategic conditions and policy priorities with little warning.

Consistent with SFA23 and FOE24, we focus on a small set of external conditions that combine high potential impact with strategic relevance. Following Sellevåg et al. (2024), we do not attempt to model every possible shock. Instead, the simulated scenario brings together several plausible developments identified in NATO foresight work: disruptive attacks on digital and economic systems; competition over scarce resources escalating to regional or global conflict; major supply-chain disruption caused by conflict or denied access to critical resources; and the emergence of a military alignment openly adversarial to NATO.

We select the hypothetical 'total trade war' scenario because it affects multiple dimensions of strategic readiness simultaneously and provides a demanding test of Europe's capacity to adjust under severe geopolitical pressure. The scenario is not a prediction. Rather, it represents an extreme but plausible condition in which geopolitical conflict extends into trade and supply networks. As Michta (2024) argues, the distinction between strategic competition and protracted conflict is already increasingly blurred. The qualitative mechanisms examined here would also apply to other forms of severe trade confrontation, including a sharp escalation of transatlantic tariffs.

To operationalise the trade-rupture scenario in EU-EMS, we impose a complete cessation of trade between the 32 NATO members plus 37 NATO partners and an 'Eastern axis' comprising China, Russia, Iran, North Korea and Belarus. All other countries are treated as neutral, with no exogenous change in trade barriers. Trade in final goods, intermediate goods and raw materials between the two blocs is eliminated by imposing prohibitively high bilateral trade costs. Trade costs within each bloc and with neutral countries remain unchanged, allowing trade patterns among unaffected partners to adjust endogenously.

We compare two policy responses to the hypothetical trade rupture: rapid and moderate trade diversion. Rapid diversion assumes an active, front-loaded effort to secure alternative markets and sources of critical inputs through multilateral, regional and bilateral engagement. Moderate diversion represents a slower adjustment path that relies more heavily on internal substitution and market-led reallocation. Comparing the two paths allows us to assess how the timing and speed of adaptation affect strategic readiness.

## 3.2 Industrial mobilisation

We first apply the stress test to industrial mobilisation, asking how a sudden loss of access to CRINK inputs would affect European defence production and how the speed of trade diversion changes the adjustment path.

For this exercise, we apply the scenario assumptions set out in Section 3.1: trade with CRINK ceases completely, and the EU-EMS model (Kancs 2024) estimates the resulting effects on defence-industrial production. The objective is not to forecast a specific event, but to compare how alternative policy trajectories alter Europe's exposure to a plausible high-impact shock.

Figure 3 reports the simulated effect of abrupt trade decoupling from CRINK on aggregate European defence production, expressed as a percentage change from the baseline. In the short run, defined as the first one to three years, defence-industrial output falls by an estimated 7.3–7.7 per cent per year. In the medium to long run, trade is progressively redirected towards Alliance members and partners, reducing the output loss to about 2.4–5.2 per cent.

**Figure 3**: European defence-industrial production (percentage change from baseline) following an abrupt decoupling from CRINK

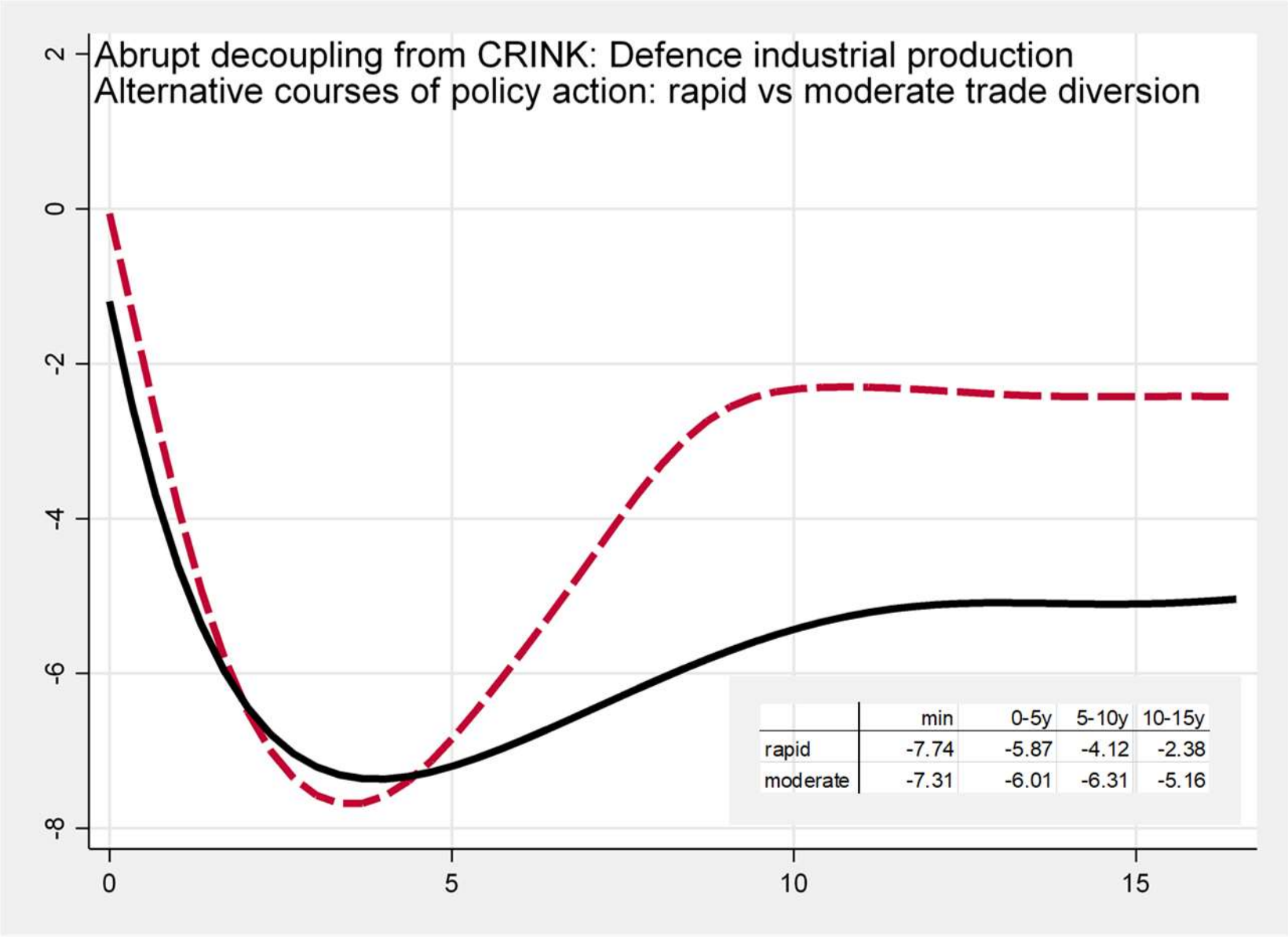

*Source: Authors' simulations based on the EU-EMS model. Notes: Values are percentage deviations from the baseline. Scenario assumptions and the rapid- and moderate-diversion policy paths are defined in Section 3.1.*

The two policy paths involve a clear short-run/long-run trade-off. Rapid trade diversion initially produces larger output losses because supply chains are reconfigured quickly towards alternative, often more expensive, sources. Moderate diversion produces smaller short-run losses by relying more heavily on internal substitution and delaying costly restructuring. Over the medium to long run, however - the horizon most relevant to a protracted conflict - rapid diversion enables a stronger recovery of defence-industrial production. The comparison therefore illustrates that front-loaded adjustment can be more costly initially but leave the industrial base better positioned once the shock persists.

In the EU-EMS model, the main effect on defence-industrial production operates through disrupted input-output linkages, particularly for intermediate goods and raw materials sourced from China (Kancs 2024). European weapon production depends on a wide range of imported inputs, including sensors for precision-guided missiles, infrared lenses for night-vision systems, nitrocellulose for propellants and high-performance fibres for body armour. China also supplies more than one-third of the raw materials used by European defence manufacturers, including large shares of rare earths, tungsten, magnesium, germanium, gallium, indium, lead and vanadium (EC 2016). These materials are embedded in critical applications such as high-performance magnets, flight-control systems and advanced microelectronics. Abrupt disruption of these inputs therefore translates directly into the negative output shock shown in Figure 3.

The defence-industrial results reinforce the supply-security findings from Section 2. A sudden loss of critical foreign inputs would reduce European defence output, while prior diversification and trade adjustment would lower the cost of the shock. The key implication is therefore one of timing: reducing concentrated dependencies before a crisis is materially less costly than being forced to adjust after supply has already been disrupted.

## 3.3 Sustained resilience

We next apply the stress test to sustained resilience. Here the focus is economic capacity: whether European economies can continue to generate the resources required for defence during a prolonged shock. Sustained military effort ultimately depends on fiscal and economic depth. Ukraine's wartime experience illustrates the scale of that requirement: defence spending has risen to around 40 per cent of GDP, supplemented by substantial foreign support (SIPRI 2025).

Because prolonged conflict depends on sustained economic and financial resources, the resilience analysis focuses on the economic capacity to absorb and recover from a major shock. Following Alloush and Carter (2024), we measure resilience as the cumulative loss experienced by a shock-exposed economy relative to a counterfactual path without the shock. In EU-EMS, the impact of the trade-rupture scenario is measured through changes in Gross National Expenditure (GNE) relative to the simulated baseline. GNE - also known as domestic absorption - is the sum of household consumption, government expenditure and investment, and is a welfare-relevant measure used in comparable macroeconomic and trade models (Baqaee et al. 2024).[11]

Figure 4 reports the effect of abrupt decoupling from CRINK on European Gross National Expenditure (GNE), measured relative to the baseline. In the first three years, GNE falls by an estimated 5.8–6.9 per cent; over five years, the loss remains about 4.4–5.7 per cent. The shock is therefore large enough to weaken the economic base on which sustained defence effort depends.

The policy comparison again highlights timing. Rapid trade diversion before the shock reduces the eventual loss relative to a slower adjustment path. This does not eliminate the cost of decoupling, but it spreads adjustment over time and reduces the disruption associated with an abrupt break. The scenario results therefore support preparedness measures that reduce concentrated exposure before geopolitical events force adjustment.

[11] GDP additionally reflects net exports and may therefore capture terms-of-trade effects differently from GNE following an extreme trade shock. Although GNE differs conceptually from GDP, the two aggregates are of a similar order of magnitude.

**Figure 4**: Gross National Expenditure in Europe (percentage change from baseline) following an abrupt decoupling from CRINK

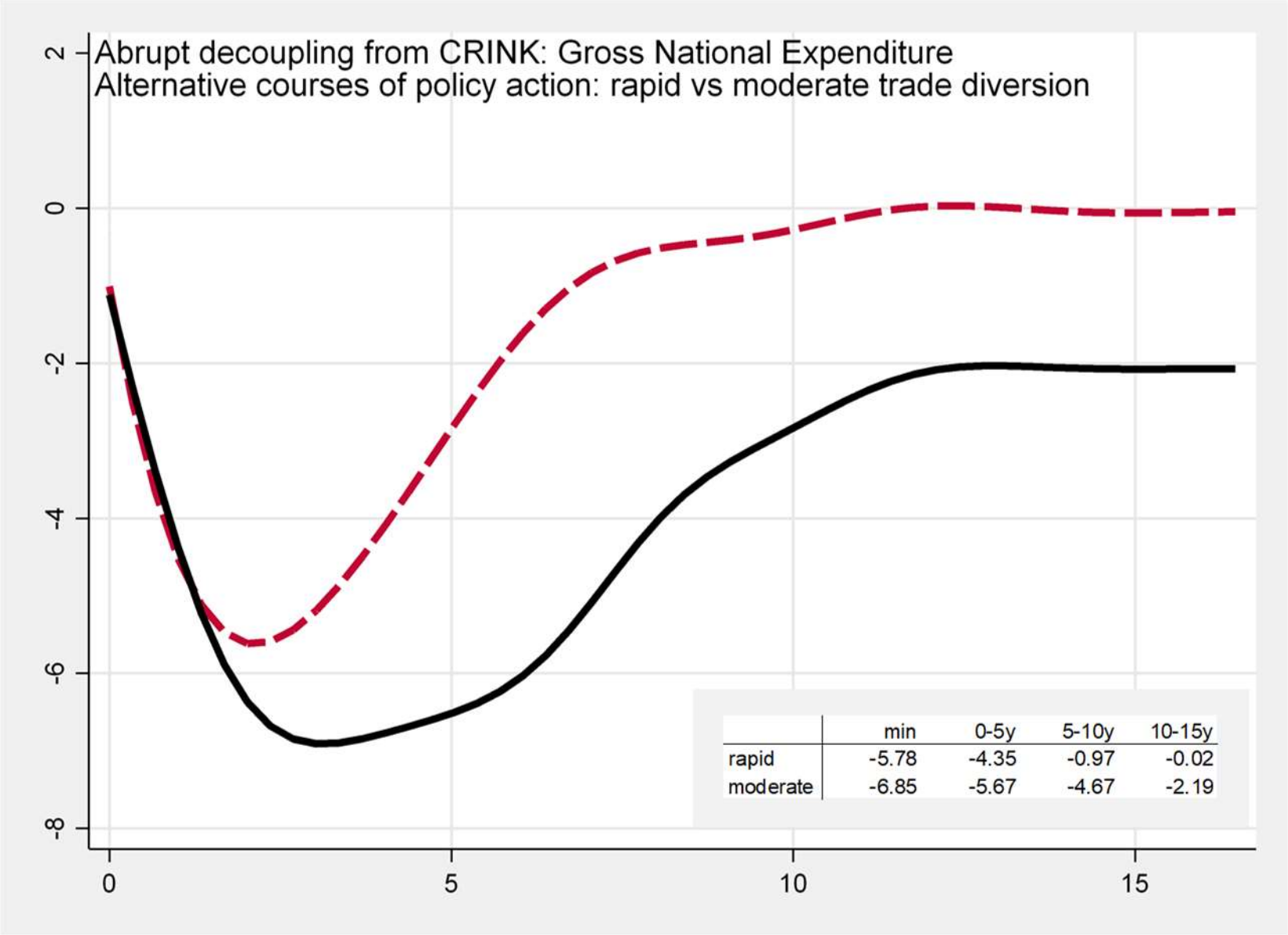

*Source: Authors' simulations based on the EU-EMS model. Notes: Values are percentage deviations from the baseline. Scenario assumptions and policy paths are defined in Section 3.1.*

These estimates should be interpreted in context. EU-EMS does not include the standard short-run business-cycle amplification mechanisms found in many macroeconomic models, which tends to make the estimates conservative in that respect. The simulated losses are somewhat larger than, but of the same order of magnitude as, those reported by Baqaee et al. (2024). In a comparable scenario involving a complete cessation of trade between Western and Eastern blocs, they estimate a short-run annual GDP loss for Germany of 5.0–5.8 per cent. Losses of this scale are severe: they are comparable to contractions experienced during the global financial crisis and the COVID-19 pandemic. Moreover, part of the effect would persist, leaving European economies on a permanently lower output path.

The scenario nevertheless suggests that losses of this magnitude can be mitigated through policy action. The results show that socio-economic costs are lower, and economic resilience stronger, when governments reduce vulnerabilities before a systemic shock occurs rather than relying on emergency adjustment afterwards. Rapid diversification away from concentrated exposure to the Eastern axis is one example of such anticipatory action.

## 3.4 Synthesis: readiness under stress

The stress test reinforces the current-readiness assessment. Foreign-input dependencies can reduce defence-industrial output at the same time as a systemic shock weakens the economic resources needed to sustain defence effort. Across both dimensions, the results point to the same conclusion: preparedness is partly a question of timing. Measures taken before a shock can preserve more capacity and lower adjustment costs than measures taken after disruption has begun.

# 4 Implications for strategic decision-makers

European Defence Readiness is ultimately a conversion problem: political commitment must be translated into investment, investment into capability and capacity, and those intermediate outputs into sustained security. The analysis identifies where this conversion process is weakest and where policy intervention can have the greatest effect.

## 4.1 Mobilisation readiness implications

The mobilisation assessment points to three priorities. First, production capacity must be sufficient to replace losses and expand stocks under prolonged demand. Second, a more integrated European defence market is needed to exploit economies of scale, reduce duplication and improve interoperability. Third, security-of-supply policy must reduce concentrated dependencies on strategic competitors and improve visibility across multi-tier supply chains. The scenario analysis adds a temporal dimension: these measures are more effective when undertaken before a crisis forces abrupt adjustment.

Europe has substantial economic and industrial resources, but mobilising them requires changes in institutions as well as spending. Nationally segmented procurement, chronic force shortages and dependence on external suppliers limit the conversion of resources into usable capability. The EU's comparative advantage lies particularly in market integration, regulation and coordination. A more coherent defence market can therefore complement national efforts by aggregating demand, supporting larger production runs and reducing avoidable duplication.

## 4.2 Whole-of-society resilience implications

Whole-of-society resilience requires a different emphasis. Nordic allies demonstrate that high levels of preparedness are achievable, but the Alliance remains exposed where national resilience is weaker. Policy should therefore focus on strengthening the prerequisites of resilience - preparedness, shock absorption, recovery and adaptation - in countries where the current-readiness assessment identifies persistent gaps. The scenario analysis also shows that economic resilience can weaken sharply under a large external shock, reinforcing the case for action before a crisis materialises.

Strengthening resilience is inherently a society-wide task. Military preparedness alone is insufficient: civilian authorities, the private sector and citizens all contribute capabilities on which military operations depend, from logistics and telecommunications to critical infrastructure and emergency response. Finland and Sweden provide useful examples of whole-of-society approaches. The policy challenge is to institutionalise these interdependencies and to prepare societies for the competition over scarce resources that a prolonged conflict would create.

# 5 Conclusions

The study reaches two principal conclusions. First, European mobilisation readiness remains constrained by shallow inventories, long industrial regeneration times, fragmented defence markets and external supply dependencies. Second, whole-of-society resilience is highly uneven across allies: Nordic countries perform strongly, while weaknesses elsewhere can reduce collective resilience. Readiness should therefore be understood as sustained capacity under stress, not simply as the availability of forces or resources at a single point in time.

The scenario analysis shows why this distinction matters. A severe trade rupture with CRINK would reduce defence-industrial output and weaken the economic resources available to sustain defence. The comparison of policy trajectories indicates that gradual de-risking and diversification before a shock can reduce these losses relative to adjustment forced by an abrupt break. The central policy implication is therefore not only what Europe should change, but also when: preparedness measures taken in advance are less costly and more effective than emergency adjustment after disruption begins.

These findings provide an analytical basis for assessing the preparedness consequences of geopolitical and security-policy choices. By linking current readiness gaps to their behaviour under stress, the study makes the cost of unpreparedness more visible and provides a measurable rationale for earlier action. Scenario analysis is most useful in this role: not as a forecast, but as a tool for comparing courses of action before decisions become urgent.

Europe's preparedness challenge is therefore structural rather than temporary. The security environment has changed faster than many of the institutions, industrial structures and resilience arrangements developed during the post-Cold War period. Strengthening readiness requires sustained improvements in mobilisation capacity, market integration, security of supply and societal resilience, rather than isolated increases in spending or short-term crisis measures.

The study has important limitations, above all the scarcity of publicly available defence-sector data. A stronger causal analysis would require detailed production data, ideally at product level, covering both input and output transactions. Such data would allow more rigorous estimation of the relationships between defence investment, industrial capacity, military capability and security outcomes. Building this evidence base remains an important priority for future research.

# 7 Appendix

## 7.1 **Appendix 1:** Table A1: Stocks of key weapon systems of selected European allies in 1990 and 2024

| 1990 | BEL | DNK | FRA | DEU | ITA | NLD | NOR | PRT | ESP | UK | FIN | SWE |
|---|---|---|---|---|---|---|---|---|---|---|---|---|
| Personnel | 85,450 | 29,400 | 453,100 | 476,300 | 361,400 | 101,400 | 32,700 | 33,100 | 257,400 | 300,100 | 27,300 | 63,000 |
| MBT | 359 | 499 | 1,349 | 7,000 | 1,220 | 913 | 211 | 146 | 838 | 1,314 | 120 | 785 |
| IFV | 667 | na | 1,965 | 5,066 | na | 984 | 53 | 166 | na | 2,201 | 72 | na |
| APC | 1,421 | 595 | 3,674 | 10,327 | 3,879 | 2,232 | 150 | 255 | 1,742 | 3,590 | 486 | 600 |
| ARTY/HOW | 604 | 946 | 3,167 | 7,328 | 3,202 | 1,287 | 402 | 148 | 2,233 | 1,441 | 636 | 1,020 |
| MOR | 433 | 548 | 1,228 | 1,274 | 1,905 | 339 | 125 | 158 | 1,665 | 500 | 1,494 | 1,500 |
| MRL | | | 11 | 556 | 2 | 22 | | | 12 | 26 | | |
| AD | 190 | 36 | 2,536 | 3,072 | 302 | 226 | | 105 | 695 | | 124 | 714 |
| SAM | 39 | | 345 | 1,047 | 271 | 474 | 108 | 12 | 50 | 120 | | |
| Submarines | 0 | 4 | 17 | 24 | 9 | 5 | 11 | 3 | 8 | 24 | 0 | 12 |
| PSC | 4 | 3 | 41 | 14 | 32 | 15 | 5 | 10 | 20 | 48 | 0 | 0 |
| Aircraft | 185 | 106 | 1,001 | 756 | 537 | 217 | 94 | 83 | 268 | 898 | 118 | 471 |
| 2024 | BEL | DNK | FRA | DEU | ITA | NLD | NOR | PRT | ESP | UK | FIN | SWE |
| Personnel | 23,500 | 13,100 | 202,200 | 179,850 | 161,850 | 33,650 | 25,400 | 21,500 | 122,200 | 141,100 | 23,850 | 14,850 |
| MBT | 0 | 44 | 200 | 313 | 150 | 0 | 36 | 34 | 274 | 213 | 200 | 110 |
| IFV | 35 | 44 | 814 | 680 | 717 | 117 | 91 | 30 | 309 | 388 | 212 | 361 |
| APC | 78 | 390 | 2,557 | 802 | 370 | 200 | 390 | 373 | 915 | 875 | 1,085 | 845 |
| ARTY/HOW | 14 | 2 | 104 | 109 | 179 | 21 | 24 | 59 | 364 | 167 | 752 | 26 |
| MOR | 46 | 15 | 132 | 98 | 508 | 101 | 143 | 234 | 1,189 | 360 | 716 | 228 |
| MRL | | 8 | 9 | 38 | 21 | 2 | | | | 26 | 75 | |
| AD | | | | | | | | | | | 407 | 30 |
| SAM | | | | | 12 | 42 | | | 54 | 50 | 60 | 20 |
| Submarines | 0 | 0 | 9 | 6 | 8 | 3 | 6 | 2 | 2 | 10 | 0 | 4 |
| PSC | 2 | 5 | 22 | 11 | 18 | 5 | 4 | 4 | 11 | 16 | 0 | 0 |
| Aircraft | 50 | 49 | 298 | 226 | 211 | 40 | 49 | 36 | 171 | 219 | 89 | 99 |

*Source: IISS (1991), Military Balance; IISS (2025), Military Balance. Notes: Active military personnel; main battle tank; infantry fighting vehicle; armoured personnel carrier; artillery gun and howitzer; mortar; multiple rocket launcher; air defence; surface-to-air missile; submarine; principal surface combatant (carrier, cruiser, destroyer and frigate); combat aircraft.*

## 7.2 Appendix 2: EU-EMS model[12]

To study how systemic shocks are transmitted through prices, production, consumption, trade and welfare in the presence of global, multi-stage production networks, we use the empirically parameterised and validated model of Kancs (2024n), adapted to capture the general-equilibrium effects of a global supply-chain shocks, and assess how changes in international trade costs are transmitted through cross-border production networks and final demand. The framework builds on the multi-sector Ricardian model of Caliendo and Parro (2015) and adopts the use-specific sourcing extension discussed by Antràs and Chor (2022). The distinguishing feature is that goods originating in the same country and sector may face different delivered costs according to whether they enter the production of a particular downstream sector or are purchased for final use. The model therefore accommodates separate bilateral sourcing patterns for intermediate inputs and final demand, while retaining a tractable input–output structure.

### 7.2.1 Economic environment

The world consists of $J$ countries, indexed by $i, j, k$, and $S$ sectors, indexed by $r, s$. The first country and sector indices identify the supplier; the second identify the purchaser. A use category $u \in \{1, \dots, S, F\}$ indicates intermediate use by sector $u = s$ or final use ($u = F$). Each country $j$ is endowed with an inelastically supplied quantity $L_j$ of equipped labour, remunerated at wage $w_j$. Firms operate under perfect competition and constant returns to scale. Labour is the only primary factor; intermediate inputs may be sourced domestically or abroad.

The representative household in country $j$ has Cobb–Douglas preferences over sectoral final-use composites:

$$U_j = \prod_{r=1}^{S} \left(C_j^r\right)^{\alpha_j^r}, \qquad \alpha_j^r \geq 0, \quad \sum_{r=1}^{S} \alpha_j^r = 1. \qquad (1)$$

A variety $\omega \in [0,1]$ of sector $s$ produced in country $j$ combines labour with composite intermediate inputs according to

$$y_j^s(\omega) = z_j^s(\omega) \left(\ell_j^s(\omega)\right)^{\beta_j^s} \prod_{r=1}^{S} \left(M_j^{rs}(\omega)\right)^{\gamma_j^{rs}}, \qquad \beta_j^s = 1 - \sum_{r=1}^{S} \gamma_j^{rs} > 0. \qquad (2)$$

Here $\gamma_j^{rs} \geq 0$ is the share of sector-$s$ gross production costs spent on inputs supplied by sector $r$, and $\beta_j^s$ is the value-added share. Sector- and country-specific productivity follows an independent Fréchet distribution,

$$\Pr(z_i^r(\omega) \leq z) = \exp\left(-T_i^r z^{-\theta^r}\right), \qquad T_i^r > 0, \quad \theta^r > \max\{1, \sigma^r - 1\}. \qquad (3)$$

$T_i^r$ measures technology, $\theta^r$ is the sector-specific trade elasticity, and $\sigma^r > 1$ is the elasticity of substitution across varieties within sector $r$. For each destination $j$, supplying sector $r$ and use $u \in \{1, \dots, S, F\}$, the composite quantity is a constant-elasticity-of-substitution (CES) aggregate of varieties purchased from the lowest delivered-price suppliers:

$$Q_j^{ru} = \left[\int_0^1 q_j^{ru}(\omega)^{(\sigma^r - 1)/\sigma^r} \, \mathrm{d}\omega\right]^{\sigma^r/(\sigma^r - 1)}, \qquad \sigma^r > 1. \qquad (4)$$

[12] https://web.jrc.ec.europa.eu/policy-model-inventory/explore/models/model-eu-ems/

The corresponding dual price index is $P_j^{ru} = \left[\int_0^1 p_j^{ru}(\omega)^{1-\sigma^r}\,\mathrm{d}\omega\right]^{1/(1-\sigma^r)}$, where $p_j^{ru}(\omega)$ is the minimum delivered price over possible origins. In the production function, $M_j^{rs} = Q_j^{rs}$; in final demand, $C_j^r = Q_j^{rF}$. Thus different purchasing sectors and final users demand the same underlying varieties through use-specific delivered-cost schedules. The CES elasticity $\sigma^r$ and the producer's Fréchet productivity draws are common across uses. This restriction isolates differences in delivered costs rather than imposing distinct preferences, technologies or elasticities for each end use.

#### ***Economic foundation of use-specific sourcing***

For goods from sector $r$ shipped from country $i$ to country $j$ for use $u$, let $\tau_{ij}^{ru} \geq 1$ denote an iceberg-equivalent delivered-cost factor, with $\tau_{jj}^{ru} = 1$. For each unit delivered, $\tau_{ij}^{ru}$ units must be produced or expended in resource-equivalent transport and fulfilment activities. Consequently, conditional on a producer's productivity draw, the delivered price is

$$p_{ij}^{ru}(\omega) = \frac{c_i^r \tau_{ij}^{ru}}{z_i^r(\omega)}. \qquad (5)$$

The difference between $\tau_{ij}^{rs}$ and $\tau_{ij}^{rF}$ reflects economically meaningful distinctions between business-to-business sourcing and final-market delivery. Intermediate purchases may require technical compatibility, specialised packaging, traceability, certification, just-in-time logistics or buyer-specific fulfilment. Final purchases may involve different distribution networks, retail preparation, product standards or delivery arrangements. Even within an aggregated input–output sector, downstream industries and households may purchase different mixes of narrowly defined products with different shipping and compliance requirements. These differences can be represented parsimoniously by effective delivered-cost wedges at the country-pair, supplying-sector and end-use level (Antràs and Chor, 2022, Section 4.1.4).

All counterfactual shocks are specified as proportional changes in iceberg-equivalent trade costs; no tariff receipts, government revenue or revenue-recycling mechanism arise. The wedges are reduced-form representations of use-specific delivery requirements, not separately modelled matching, fixed sourcing costs or contractual incompleteness. When interpreting results note that observed differences in end-use trade shares may also reflect product composition or input–output table construction, rather than distinct structural frictions. The baseline shares alone cannot distinguish among those explanations.

### 7.2.2 Prices and sourcing decisions

Cost minimisation under (2) gives the unit cost before the productivity draw:

$$c_j^s = \kappa_j^s w_j^{\beta_j^s} \prod_{r=1}^{S} \left(P_j^{rs}\right)^{\gamma_j^{rs}}, \qquad \kappa_j^s = \left(\beta_j^s\right)^{-\beta_j^s} \prod_{r:\gamma_j^{rs}>0} \left(\gamma_j^{rs}\right)^{-\gamma_j^{rs}}. \qquad (6)$$

$P_j^{rs}$ is the price of the sector-$r$ composite used by sector $s$ in country $j$; $P_j^{rF}$ denotes the corresponding final-use price. For each variety and use, the purchaser selects the lowest delivered-price country. Define $\Phi_j^{ru} = \sum_{k=1}^{J} T_k^r \left(c_k^r \tau_{kj}^{ru}\right)^{-\theta^r}$. The sector-use price index and the share of country $j$'s spending on that composite obtained from source $i$ are

$$P_j^{ru} = A^r \left(\Phi_j^{ru}\right)^{-1/\theta^r}, \qquad A^r = \Gamma(1 - (\sigma^r - 1)/\theta^r)^{1/(1-\sigma^r)}.$$

$$\pi_{ij}^{ru} = \frac{T_i^r \left(c_i^r \tau_{ij}^{ru}\right)^{-\theta^r}}{\sum_{k=1}^{J} T_k^r \left(c_k^r \tau_{kj}^{ru}\right)^{-\theta^r}}. \qquad (7)$$

For every active destination–use market, $\sum_i \pi_{ij}^{ru} = 1$. $\Gamma(\cdot)$ is the gamma function. Under the restriction in (3), $A^r$ is finite, does not vary with the purchaser or use, and cancels from exact-hat counterfactuals. The same productivity and producer-cost terms enter every use of sector $r$; only the delivered-cost factors differ. Consequently, sectors and final consumers can have different bilateral sourcing shares, even when their underlying variety aggregators are identical. An increase in $\tau_{ij}^{ru}$ directly reduces the relative attractiveness of origin $i$ for that particular use. For $i \neq k$, the elasticity of relative sourcing odds with respect to $\tau_{ij}^{ru}$ is $-\theta^r$, holding other costs constant. Equilibrium wages and input prices also respond, so the total change in trade is not simply this partial-equilibrium elasticity.

If trade costs are common across uses, $\tau_{ij}^{ru} = \tau_{ij}^r$, equation (7) implies $\pi_{ij}^{rs} = \pi_{ij}^{rF}$ for every purchasing sector $s$. This proportionality restriction characterises the benchmark Caliendo–Parro setup and is often at odds with detailed input–output data (Antràs and Chor, 2022, Section 4.1.2). Allowing $\tau_{ij}^{ru}$ to vary with use relaxes it without changing production technology or the sector-specific Fréchet elasticity.

The aggregate final-consumption price index is distinct from the sector-use prices:

$$P_j^C = \prod_{r:\alpha_j^r>0} \left(\frac{P_j^{rF}}{\alpha_j^r}\right)^{\alpha_j^r}. \qquad (8)$$

### 7.2.3 Market clearing and counterfactual equilibrium

Let $Y_j^s$ denote gross output valued at producer receipts. Production in country $j$, sector $s$, entails expenditure $\gamma_j^{rs} Y_j^s$ on inputs supplied by sector $r$. In the absence of tariff revenue, profits or other factor income, household expenditure is financed by labour income and an exogenous external balance $D_j$:

$$I_j = w_j L_j + D_j, \qquad E_j^{rs} = \gamma_j^{rs} Y_j^s, \qquad E_j^{rF} = \alpha_j^r I_j, \qquad \sum_j D_j = 0. \qquad (9)$$

A positive $D_j$ denotes a trade deficit. Gross-output market clearing and the national value-added identity require

$$Y_i^r = \sum_{j=1}^{J} \left( \sum_{s=1}^{S} \pi_{ij}^{rs} \gamma_j^{rs} Y_j^s + \pi_{ij}^{rF} \alpha_j^r I_j \right). \qquad (10)$$

The country-level value-added identity is $w_j L_j = \sum_{s=1}^{S} \beta_j^s Y_j^s$.

Together, (9)–(10) imply the country-level external-balance identities: expenditure on foreign intermediate and final goods less foreign expenditure on domestically supplied goods equals $D_j$. One wage is normalised as the numeraire and the global balance constraint removes one redundant national condition. External balances are not freely adjustable residuals: their counterfactual closure must be stated in advance. The reference specification holds $D_j$ fixed in units of the chosen numeraire, and sensitivity exercises consider alternative balanced closures.

For any positive variable $x$, a hat indicates the ratio between counterfactual and benchmark values, $\hat{x} = x'/x$. A counterfactual experiment is defined by the proportional shocks $\hat{\tau}_{ij}^{ru}$, with $\hat{\tau}_{jj}^{ru} = 1$.

Sectoral technology, factor endowments, input coefficients and final-expenditure shares are held constant; primes denote counterfactual levels. The exact-hat system governing costs and sourcing is

$$\hat{c}_j^s = \left(\hat{w}_j\right)^{\beta_j^s} \prod_{r=1}^{S} \left(\hat{P}_j^{rs}\right)^{\gamma_j^{rs}}, \qquad (11)$$

$$H_j^{ru} = \sum_{i=1}^{J} \pi_{ij}^{ru} \left(\hat{c}_i^r \hat{\tau}_{ij}^{ru}\right)^{-\theta^r}, \qquad \hat{P}_j^{ru} = \left(H_j^{ru}\right)^{-1/\theta^r}, \qquad (12)$$

$$\pi_{ij}^{ru\prime} = \frac{\pi_{ij}^{ru}\left(\hat{c}_i^r \hat{\tau}_{ij}^{ru}\right)^{-\theta^r}}{H_j^{ru}}. \qquad (13)$$

To close the counterfactual economy, define a $JS \times JS$ input coefficient matrix $\mathbf{M}'$ and a $JS \times 1$ final-demand vector $\mathbf{f}'$. Rows of $\mathbf{M}'$ are indexed by the producing country-sector $(i, r)$, and columns by the purchasing country-sector $(j, s)$. Their entries and the final-demand vector are

$$\begin{aligned} M'_{(i,r),(j,s)} &= \pi_{ij}^{rs\prime} \gamma_j^{rs}, \\ f'_{(i,r)} &= \sum_{j=1}^{J} \pi_{ij}^{rF\prime}\, \alpha_j^r I'_j, \\ I'_j = \hat{w}_j V_j + D'_j, &\quad V_j = \sum_s V_j^s = w_j L_j. \end{aligned} \qquad (14)$$

Conditional on wages and sourcing shares, the post-shock gross-output vector is the unique solution of the goods-market system

$$\begin{aligned} \mathbf{Y}' &= \mathbf{M}'\mathbf{Y}' + \mathbf{f}', \\ \mathbf{Y}' &= \left(\mathbf{I}_{JS} - \mathbf{M}'\right)^{-1}\mathbf{f}'. \end{aligned} \qquad (15)$$

The inverse exists and is non-negative under the maintained non-negative expenditure coefficients and strictly positive sectoral value-added shares, which imply $\sum_{i,r} M'_{(i,r),(j,s)} = \sum_r \gamma_j^{rs} < 1$ for each purchasing country-sector. For all countries, labour-market clearing and national value-added consistency require

$$\begin{aligned} \hat{w}_j V_j &= \sum_{s=1}^{S} \beta_j^s\, Y_j^{s\prime}, \\ D'_j = D_j, &\quad \sum_j D_j = \sum_j D'_j = 0, \\ \hat{w}_{\text{USA}} &= 1. \end{aligned} \qquad (16)$$

Only baseline national wage bills $V_j$, rather than separate observations on $w_j$ and $L_j$, are needed for the hat system. The United States wage is the numeraire in the reference implementation - a standard price normalisation. External balances are fixed at their benchmark nominal levels expressed in units of this numeraire. The $J$ national value-added identities contain one redundant equation by the global budget identity, leaving $J - 1$ independent relative-wage conditions. Equations (11)–(16), the calibrated benchmark objects $\{\pi, \gamma, \beta, \alpha, wL, D\}$ and the imposed $\hat{\tau}$ therefore define a closed nonlinear system for relative wages, unit-cost hats, use-specific price hats, sourcing shares and gross output. The notation $\mathbf{I}_{JS}$ denotes the $JS \times JS$ identity matrix, which is distinct from final expenditure $I_j$.

Counterfactual bilateral intermediate and final-use flows follow from the expenditure identities, while real final expenditure measures welfare in this representative-agent specification:

$$Z_{ij}^{rs\prime} = \pi_{ij}^{rs\prime}\gamma_j^{rs}Y_j^{s\prime}, \qquad F_{ij}^{r\prime} = \pi_{ij}^{rF}\ \alpha_j^r I'_j,$$
$$\widehat{W}_j = \frac{I'_j/I_j}{\hat{P}_j^C}, \qquad \hat{P}_j^C = \prod_r \left(\hat{P}_j^{rF}\right)^{\alpha_j^r}. \qquad (17)$$

These equations also supply a direct numerical verification: recomputed intermediate and final bilateral flows must sum to counterfactual sectoral output, and counterfactual value added must equal wage income up to numerical tolerance.

***Distinguishing transmission channels***

Two mechanisms link a trade-cost shock to domestic outcomes. First, intermediate-input propagation operates through $\hat{\tau}_{ij}^{rs}$, input sourcing shares, purchasing-sector input price indices and the unit costs of downstream producers. The effects are amplified or attenuated by input–output linkages as changes in costs affect the competitiveness and sourcing of subsequent stages. Second, final-demand reallocation operates directly through $\hat{\tau}_{ij}^{rF}$, changing the composition and price of household purchases and therefore demand for producing sectors. Both mechanisms trigger general-equilibrium responses in wages, production and trade shares; a shock entering through only one use category can subsequently affect both intermediate and final transactions.

The empirical design therefore produces three counterfactuals for any common shock scenario: (i) an *intermediate-only* experiment, in which $\hat{\tau}_{ij}^{rs}$ changes while $\hat{\tau}_{ij}^{rF} = 1$; (ii) a *final-only* experiment, with $\hat{\tau}_{ij}^{rs} = 1$ and the designated final-use shocks applied; and (iii) a *joint* experiment, with both blocks shocked. Results for each experiment include gross output, input costs, sector-use sourcing shares, bilateral flows, wages and real expenditure. The difference between the joint result and the sum of the two separate changes measures a general-equilibrium interaction. This scenario comparison attributes the point of entry of the shocks, rather than providing a unique additive decomposition of their subsequent propagation. For any strictly positive outcome $X$ (for example, real expenditure or gross output), the interaction is reported in log points as

$$\mathcal{J}_X = \log\left(X^{\text{joint}}/X^0\right) - \log\left(X^{\text{intermediate}}/X^0\right) - \log\left(X^{\text{final}}/X^0\right). \qquad (18)$$

A non-zero $\mathcal{J}_X$ captures nonlinear general-equilibrium interactions. For variables that may be zero or negative, including bilateral trade balances, we report level changes instead without applying logarithms. This interaction is a comparison of scenario outcomes.

### 7.2.4 Empirical implementation

***Data construction and benchmark calibration***

For a reproducible reference implementation, the baseline model uses the OECD ICIO 2025 edition (2026 revision) in current prices. The regular table covers 80 economies and a rest-of-world aggregate across 50 industries. The post-pandemic year is selected to avoid using the extraordinary 2020 disruption as the benchmark; 2018 and 2022 are robustness reference years. The standard country-industry classification is retained, including the rest-of-world region and services; the extended China/Mexico firm-split tables are not mixed into the benchmark. Results from different reference years are solved as separate benchmarks rather than combining nominal flows across years (OECD, 2026). Let $Z_{ij}^{rs}$ be the value of sector-$r$ output from country $i$ purchased as an intermediate by sector $s$ in country $j$; let $F_{ij}^{r}$ denote final-use purchases in $j$ of sector-$r$ output produced in $i$. The dataset must also contain gross output $Y_j^s$ and value added $V_j^s$. Domestic transactions must be retained. Valuation concepts must be reconciled, and discrepancies in intermediate use, output, value added, re-exports and final-use categories documented before calibration.

Conditional on positive denominators, the technology coefficients and intermediate sourcing shares are calculated as

$$\gamma_j^{rs} = \frac{\sum_i Z_{ij}^{rs}}{Y_j^s}, \qquad \beta_j^s = \frac{V_j^s}{Y_j^s}, \qquad \pi_{ij}^{rs} = \frac{Z_{ij}^{rs}}{\sum_k Z_{kj}^{rs}}. \qquad (19)$$

The corresponding final-demand coefficients and sourcing shares are

$$\alpha_j^r = \frac{\sum_i F_{ij}^r}{\sum_{q=1}^S \sum_i F_{ij}^q}, \qquad \pi_{ij}^{rF} = \frac{F_{ij}^r}{\sum_k F_{kj}^r}. \qquad (20)$$

Where the accounting boundaries coincide, $w_j L_j = \sum_s V_j^s$ and $D_j = \sum_{i,r} F_{ij}^r - \sum_s V_j^s$. The calibration is valid only if $\beta_j^s + \sum_r \gamma_j^{rs} = 1$ and the row and column identities are reconciled within stated tolerances. Recorded value added may include capital income and taxes net of subsidies; identifying it with payments to the model's single equipped-labour factor is an effective-factor normalisation, rather than a literal measure of employee compensation. Treating all WIOT final use as household expenditure is an aggregation assumption: it includes government consumption and investment insofar as they appear in the observed final-use block, and the associated welfare result is interpreted as real aggregate final expenditure rather than a full measure of household utility under disaggregated final uses. Undefined shares in zero-expenditure cells are not imputed as if observed; the relevant cell is aggregated, excluded when immaterial, or treated under a documented regularisation rule.

Equations (19)–(20) calibrate the observed benchmark spending patterns, but do not separately identify the primitives $T_i^r$, $c_i^r$ and $\tau_{ij}^{ru}$. In particular, use-specific differences in observed sourcing odds are consistent with relative differences in iceberg-equivalent costs conditional on the common-productivity model. With domestic trade costs normalised to one, the structural implication for two uses $u$ and $v$ is

$$\log \frac{\pi_{ij}^{ru} / \pi_{jj}^{ru}}{\pi_{ij}^{rv} / \pi_{jj}^{rv}} = -\theta^r \log \frac{\tau_{ij}^{ru}}{\tau_{ij}^{rv}}, \qquad i \neq j. \qquad (21)$$

Thus the baseline shares can discipline *relative* use-specific cost gaps given $\theta^r$ and the maintained assumptions, but not absolute trade costs or their structural origins. This is not an independent test of the use-specific sourcing mechanism. Moreover, some WIOT construction procedures impose proportionality between final and intermediate imports, which can mechanically suppress precisely the differences this extension is intended to represent. Data-source and concordance sensitivity is therefore integral to the analysis.

### *External parameters, trade-cost scenarios and computation*

The remaining inputs are sector-specific trade elasticities $\theta^r$ and exogenous proportional changes $\hat{\tau}_{ij}^{ru}$. The reference calculation sets $\theta^r = 4$ and $\sigma^r = 2$ for all supplying industries, with $\theta^r = 2$ and 8 as low- and high-trade-elasticity checks (Figure A1). The choice of $\sigma^r = 2$ ensures the CES price index is finite under each of these elasticity assumptions but does not otherwise affect exact-hat results. For robustness, we also test sectoral calibration based on econometric estimates (Caliendo and Parro 2015; Imbs and Méjean 2017). The supplying sector's elasticity is held constant across purchasing uses. Estimates discussed by Caliendo and Parro (2015) and Imbs and Méjean (2017) provide foundation for the heterogeneous-elasticity specification.

The trade-rupture scenario imposes a 100% rise in all *international* iceberg-equivalent delivered-cost factors, $\hat{\tau}_{ij}^{ru} = 2.0$ for $i \neq j$, leaving domestic costs unchanged. For robustness, three separate experiments impose this increase on intermediate uses only ($u = s$), final use only ($u = F$), and both, as defined above. Additional stress tests use 20% increases and decreases and, where warranted, restrict the shock to specified origin–destination pairs or supplying sectors. Note that in the underlying Ricardian framework, finite iceberg trade costs do not generate an exact zero bilateral trade share. A complete trade severance could therefore be represented either by an explicit bilateral trade prohibition or, for numerical approximation, by a sufficiently large trade-cost multiplier such that the resulting sourcing share is economically negligible

The numerical solution proceeds in proportional changes. Given trial wages, solve (11)–(12) jointly for unit-cost and price hats, calculate (13), and form the post-shock input matrix and final-demand vector in (14). We solve the sparse linear system in (15) for gross output and adjust the $J - 1$ free wages until the corresponding conditions in (16) hold. The US wage remains the numeraire, $D'_j = D_j$, and all results follow from (17). Iteration continues until the maximum absolute relative error in market-clearing and value-added identities is below $10^{-8}$ and all active sourcing shares sum to one within $10^{-10}$; these are numerical tolerances that are reported alongside convergence diagnostics. The unshocked experiment, $\hat{\tau}_{ij}^{ru} = 1$ for every cell, reproduces all benchmark flows and unit hats to the stated tolerance. Numeraire invariance is checked by rescaling wages, prices and the nominal deficit vector consistently, with real allocations unchanged.

### ***Accounting, validation and uncertainty***

Calibration and internal consistency. Reproducing observed bilateral flows, expenditure shares, gross output and the aggregate national accounts is a requirement of the calibrated benchmark, not an external validation result. The computation is checked against input–output row and column totals, positivity conditions, shares summing to one for active cells, market-clearing residuals, and the zero-shock replication exercise. The model's exact baseline fit is obtained by construction once the requisite degrees of freedom have been calibrated; it offers no independent evidence that counterfactual substitution or propagation is correctly specified (Antràs and Chor 2022, Section 4.1.5).

External validation: Detailed out-of-sample exercise and causal validation results are presented in Kancs (2024). Once the calibration and counterfactuals are implemented, an independent diagnostic uses a pre-shock benchmark year, an externally measured and empirically estimated trade-cost change and subsequent changes in trade flows, sectoral production and prices that were *not* used to choose elasticities or shock magnitudes. The proportion of correctly predicted directional movements, median absolute log prediction errors for positive flows, and discrepancies in country-sector output changes are reported; the sample restrictions and treatment of zeros is detailed. Our comparisons acknowledge simultaneous demand, technology, exchange-rate and policy shocks and cannot be interpreted causally without an identification strategy.

Uncertainty and limitations. Sensitivity analysis varies sectoral elasticities, the split between intermediate and final shocks, the baseline year and data provider, the treatment of zero and small flows, rest-of-world aggregation and the external-balance closure and wage numeraire. In exact-hat algebra, a zero baseline source share remains zero under any finite counterfactual multiplier, making network sparsity and the treatment of unobserved trade links substantively important. The underlying framework is comparative-static: it describes changes between equilibria. Introducing model adjustment time, supplier-search costs, inventories, unemployment, household heterogeneity or nominal rigidities offer a promising avenue for future research. Scenarios labelled short- and long-run on the basis of different elasticity vectors are consequently alternative comparative-static assumptions, not simulated transition paths.

The main results are reported separately for the intermediate-only, final-only and joint scenarios, including changes in bilateral trade, country-sector production, input and consumer prices, domestic value added, nominal wages and real final expenditure. These outcomes together allow to identify how the same resource-cost shock propagates through international production networks and final markets, while distinguishing evidence-based calibration from model-dependent welfare predictions.

**Figure A1:** Estimates of the elasticity of substitution among traded goods from different countries of origin over different time horizons.

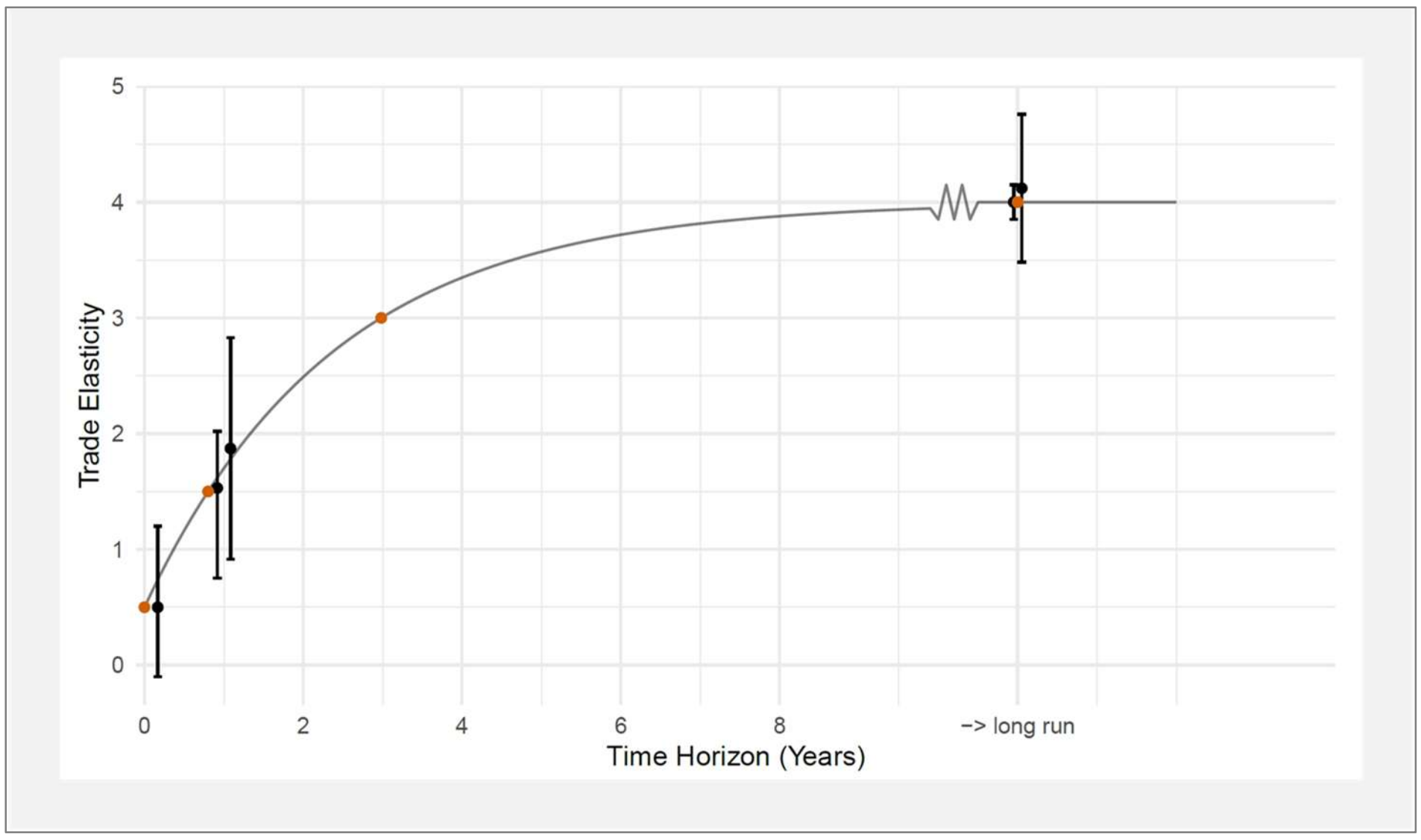

Source: Based on Baqaee et al. (2024). Notes: Higher elasticities imply easier substitution across countries of origin; see Appendix 2 for the role of trade elasticities in the model.

## List of abbreviations and definitions

| | |
|---|---|
| Agile Combat Employment | ACE |
| Allied Command Transformation | ACT |
| Armoured Personnel Carriers | APC |
| Artillery (guns, towed and self-propelled howitzers) | ARTY/HOW |
| China, Russia, Iran and North Korea | CRINK |
| Combined Nomenclature | CN |
| Constant Elasticity of Substitution | CES |
| EU Economic Modelling System | EU-EMS |
| Enhanced Analytic Resilience Index | EARI |
| Euro-Atlantic Resilience Centre | E-ARC |
| European Defence Fund | EDF |
| European Defence Industrial Strategy | EDIS |
| Eurostat External Trade Statistics | COMEXT |
| European Peace Facility | EPF |
| European Union | EU |
| Foreign Input Reliance | FIR |
| Foreign Market Reliance | FMR |
| Future Operating Environment 2024 | FOE24 |
| Herfindahl-Hirschman Index | HHI |
| Gross National Expenditure | GNE |
| International Concept Development & Experimentation | ICD&E |
| Industrial Capacity Expansion Pledge | ICEP |
| Infantry Fighting Vehicles | IFV |
| Inter-Country Input–Output | ICIO |
| International Monetary Fund | IMF |
| International Institute for Strategic Studies | IISS |
| Joint Research Centre | JRC |
| Main Battle Tanks | MBT |
| Military Instrument of Power | MIoP |
| Military Planning and Conduct Capability | MPCC |
| Military Schengen agreement | MSA |
| Mortars | MOR |
| Statistical Classification of Economic Activities | NACE |
| North Atlantic Treaty Organization | NATO |
| Operations Research and Analysis | OR&A |
| Organisation for Economic Co-operation and Development | OECD |
| Principal Surface Combatant | PSC |
| Research & Development | R&D |
| Science & Technology Organisation of NATO | STO |
| Stockholm Peace Research Institute | SIPRI |
| Strategic Foresight Analysis 2023 | SFA23 |
| United Nations Framework Convention on Climate Change | UNFCCC |
| United States Department of Defence | US DoD |
| Whiteshield | WS |
| World Input-Output Tables | WIOT |

## List of tables

## List of figures

## Getting in touch with the EU

### In person

All over the European Union there are hundreds of Europe Direct centres. You can find the address of the centre nearest you online (european-union.europa.eu/contact-eu/meet-us_en).

### On the phone or in writing

Europe Direct is a service that answers your questions about the European Union. You can contact this service:

— by freephone: 00 800 6 7 8 9 10 11 (certain operators may charge for these calls),

— at the following standard number: +32 22999696,

— via the following form: european-union.europa.eu/contact-eu/write-us_en.

## Finding information about the EU

### Online

Information about the European Union in all the official languages of the EU is available on the Europa website (european-union.europa.eu).

### EU publications

You can view or order EU publications at op.europa.eu/en/publications. Multiple copies of free publications can be obtained by contacting Europe Direct or your local documentation centre (european-union.europa.eu/contact-eu/meet-us_en).

### EU law and related documents

For access to legal information from the EU, including all EU law since 1951 in all the official language versions, go to EUR-Lex (eur-lex.europa.eu).

### EU open data

The portal data.europa.eu provides access to open datasets from the EU institutions, bodies and agencies. These can be downloaded and reused for free, for both commercial and non-commercial purposes. The portal also provides access to a wealth of datasets from European countries.

Science for policy

The Joint Research Centre (JRC) provides independent, evidence-based knowledge and science, supporting EU policies to positively impact society

**EU Science Hub**
Joint-research-centre.ec.europa.eu

Publications Office of the European Union